\documentclass[aps,pra,twocolumn,showpacs,notitlepage,superscriptaddress,letterpaper]{revtex4-1}

\usepackage{graphicx, amsmath, amssymb, amsfonts, pifont, bm, cancel, bbold}
\usepackage[usenames]{color}
\usepackage{subfigure}
\usepackage[normalem]{ulem} 
\usepackage{hyperref}
\usepackage{threeparttable}
\usepackage{array}
\newcolumntype{P}[1]{>{\centering\arraybackslash}p{#1}}
\newcolumntype{M}[1]{>{\centering\arraybackslash}m{#1}}

\definecolor{MyDarkBlue}{rgb}{0,0,1}

\newcommand{\bel}{\begin{align*}}
\newcommand{\tamam}{\end{align*}}

\newcommand{\sigmax}{\hat{\sigma}_x}

\newcommand{\sigmaz}{\hat{\sigma}_z}

\newcommand{\QL}{Q$_\text{L}$} 
\newcommand{\QR}{Q$_\text{R}$} 
\newcommand{\EL}{E$_\text{L}$} 
\newcommand{\ER}{E$_\text{R}$} 
\newcommand{\gL}{$g_\text{L}$} 
\newcommand{\gR}{$g_\text{R}$} 

\newcommand{\ud}{\mathrm{d}}

\begin{document}

\title{Quantum electrodynamics in a topological waveguide}
\author{Eunjong~Kim}
\thanks{These authors contributed equally to this work.}
\author{Xueyue~Zhang}
\thanks{These authors contributed equally to this work.}
\author{Vinicius~S.~Ferreira}
\author{Jash Banker}
\affiliation{Thomas J. Watson, Sr., Laboratory of Applied Physics and Kavli Nanoscience Institute, California Institute of Technology, Pasadena, California 91125, USA.}
\affiliation{Institute for Quantum Information and Matter, California Institute of Technology, Pasadena, California 91125, USA.}
\author{Joseph K.~Iverson}
\affiliation{Institute for Quantum Information and Matter, California Institute of Technology, Pasadena, California 91125, USA.}
\affiliation{AWS Center for Quantum Computing, Pasadena, California 91125, USA.}
\author{Alp Sipahigil}
\affiliation{Thomas J. Watson, Sr., Laboratory of Applied Physics and Kavli Nanoscience Institute, California Institute of Technology, Pasadena, California 91125, USA.}
\affiliation{Institute for Quantum Information and Matter, California Institute of Technology, Pasadena, California 91125, USA.}
\author{Miguel Bello}
\affiliation{Instituto de Ciencia de Materiales de Madrid, CSIC, 28049 Madrid, Spain.}
\author{Alejandro Gonz\'{a}lez-Tudela}
\affiliation{Instituto de F\'{i}sica Fundamental, IFF-CSIC, Calle Serrano 113b, Madrid 28006, Spain.}
\author{Mohammad~Mirhosseini}
\affiliation{Institute for Quantum Information and Matter, California Institute of Technology, Pasadena, California 91125, USA.}
\affiliation{Gordon and Betty Moore Laboratory of Engineering, California Institute of Technology, Pasadena, California 91125, USA.}
\author{Oskar~Painter}
\email{opainter@caltech.edu}
\homepage{http://copilot.caltech.edu}
\affiliation{Thomas J. Watson, Sr., Laboratory of Applied Physics and Kavli Nanoscience Institute, California Institute of Technology, Pasadena, California 91125, USA.}
\affiliation{Institute for Quantum Information and Matter, California Institute of Technology, Pasadena, California 91125, USA.}
\affiliation{AWS Center for Quantum Computing, Pasadena, California 91125, USA.}

\date{\today}


\begin{abstract} 

While designing the energy-momentum relation of photons is key to many linear, non-linear, and quantum optical phenomena, a new set of light-matter properties may be realized by employing the topology of the photonic bath itself. In this work we investigate the properties of superconducting qubits coupled to a metamaterial waveguide based on a photonic analog of the Su-Schrieffer-Heeger model. We explore topologically-induced properties of qubits coupled to such a waveguide, ranging from the formation of directional qubit-photon bound states to topology-dependent cooperative radiation effects. Addition of qubits to this waveguide system also enables direct quantum control over topological edge states that form in finite waveguide systems, useful for instance in constructing a topologically protected quantum communication channel. More broadly, our work demonstrates the opportunity that topological waveguide-QED systems offer in the synthesis and study of many-body states with exotic long-range quantum correlations.
\end{abstract}

\maketitle

Harnessing the topological properties of photonic bands~\cite{Haldane:2008,Lu:2014,Ozawa:2019} is a burgeoning paradigm in the study of periodic electromagnetic structures. Topological concepts discovered in electronic systems~\cite{vonKlitzing:1986,Hasan:2010} have now been translated and studied as photonic analogs in various microwave and optical systems~\cite{Lu:2014,Ozawa:2019}. In particular, symmetry-protected topological phases~\cite{Chen:2013} which do not require time-reversal-symmetry breaking, have received significant attention in experimental studies of photonic topological phenomena, both in the linear and nonlinear regime~\cite{Smirnova:2019}. One of the simplest canonical models is the Su-Schrieffer-Heeger (SSH) model~\cite{Su:1979, Asboth:2016}, which was initially used to describe electrons hopping along a one-dimensional dimerized chain with a staggered set of hopping amplitudes between nearest-neighbor elements. The chiral symmetry of the SSH model, corresponding to a symmetry of the electron amplitudes found on the two types of sites in the dimer chain, gives rise to two topologically distinct phases of electron propagation.  The SSH model, and its various extensions, have been used in photonics to explore a variety of optical phenomena, from robust lasing in arrays of microcavities~\cite{St-Jean:2017,Zhao:2018} and photonic crystals~\cite{Ota:2018}, to disorder-insensitive 3rd harmonic generation in zigzag nanoparticle arrays~\cite{Kruk:2019}. 

Utilization of quantum emitters brings new opportunities in the study of topological physics with strongly interacting photons~\cite{Carusotto:2013}, where single-excitation dynamics~\cite{Cai:2019} and topological protection of quantum many-body states~\cite{deLeseleuc:2019} in the SSH model have recently been investigated.  In a similar vain, a topological photonic bath can also be used as an effective substrate for endowing special properties to quantum matter. For example, a photonic waveguide which localizes and transports electromagnetic waves over large distances, can form a highly effective quantum light-matter interface~\cite{Haroche:2006,Lodahl:2015,Chang:2018} for introducing non-trivial interactions between quantum emitters.  Several systems utilizing highly dispersive electromagnetic waveguide structures have been proposed for realizing quantum photonic matter exhibiting tailorable, long-range interactions between quantum emitters~\cite{Douglas:2015,Shi:2018,Liu:2016,Hung:2016}.  With the addition of non-trivial topology to such a photonic bath, exotic classes of quantum entanglement can be generated through photon-mediated interactions of a chiral~\cite{Barik:2018, Lodahl:2017} or directional nature~\cite{Bello:2019,Garcia-Elcano:2019}.

With this motivation, here we investigate the properties of quantum emitters coupled to a topological waveguide which is a photonic analog of the SSH model~\cite{Bello:2019}. Our setup is realized by coupling superconducting transmon qubits~\cite{Koch:2007} to an engineered superconducting metamaterial waveguide~\cite{Mirhosseini:2018,Ferreira:2020}, consisting of an array of sub-wavelength microwave resonators with SSH topology. Combining the notions from waveguide quantum electrodynamics (QED)~\cite{Lodahl:2015, Chang:2018, Gu:2019, Roy:2017} and topological photonics~\cite{Lu:2014, Ozawa:2019}, we observe qubit-photon bound states with directional photonic envelopes inside a bandgap and cooperative radiative emission from qubits inside a passband dependent on the topological configuration of the waveguide.  Coupling of qubits to the waveguide also allows for quantum control over topological edge states, enabling quantum state transfer between distant qubits via a topological channel.

\begin{figure*}[t!]
\begin{center}
\includegraphics[width=1\textwidth]{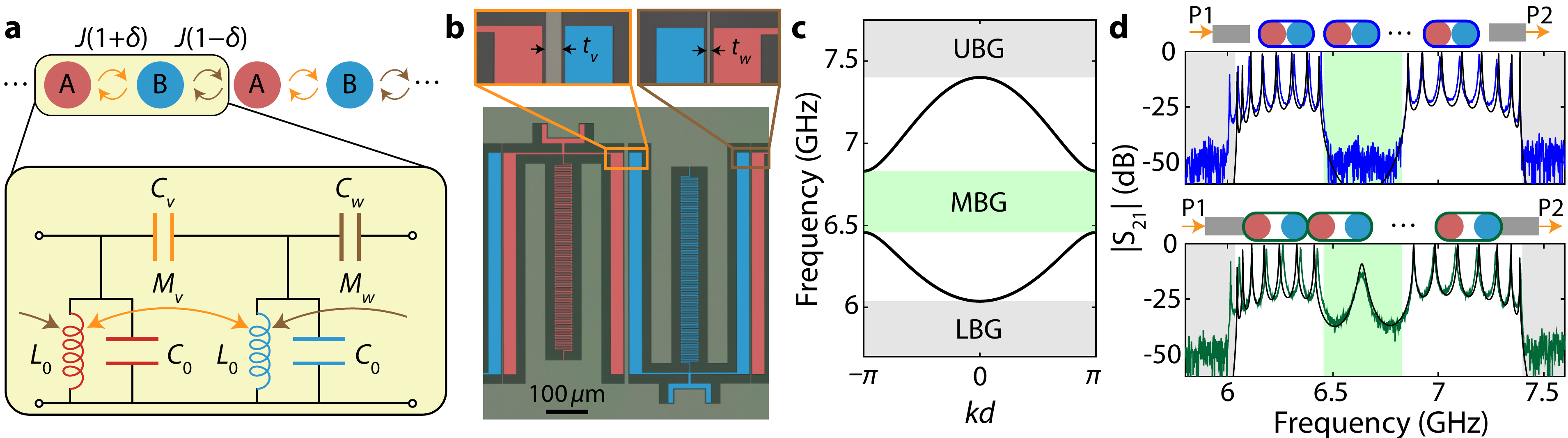}
\caption{\textbf{Topological waveguide.} \textbf{a}, Top: schematic of the SSH model. Each unit cell contains two sites A and B (red and blue circles) with intra- and inter-cell coupling $J(1\pm\delta)$ (orange and brown arrows). Bottom: an analog of this model in electrical circuits, with corresponding components color-coded. The photonic sites are mapped to LC resonators with inductance $L_0$ and capacitance $C_0$, with intra- and inter-cell coupling capacitance $C_v$, $C_w$ and mutual inductance $M_v$, $M_w$ between neighboring resonators, respectively (arrows). \textbf{b}, Optical micrograph (false-colored) of a unit cell (lattice constant $d=592\:\mu\text{m}$) on a fabricated device in the topological phase. The lumped-element resonator corresponding to sublattice A (B) is colored in red (blue). The insets show zoomed-in view of the section between resonators where planar wires of thickness $(t_v,t_w)=(10,2)\:\mu\text{m}$ (indicated with black arrows) control the intra- and inter-cell distance between neighboring resonators, respectively. \textbf{c}, Dispersion relation of the realized waveguide according to the circuit model in panel \textbf{a}. Upper bandgap (UBG) and lower bandgap (LBG) are shaded in gray, and middle bandgap (MBG) is shaded in green. \textbf{d}, Waveguide transmission spectrum $|S_{21}|$ across the test structure with 8 unit cells in the trivial ($\delta > 0$; top) and topological ($\delta < 0$; bottom) phase. The cartoons illustrate the measurement configuration of systems with external ports 1 and 2 (denoted P1 and P2), where distances between circles are used to specify relative coupling strengths between sites and blue (green) outlines enclosing two circles indicate unit cells in the trivial (topological) phase. Black solid curves are fits to the measured data (see App.~\ref{sec:AppendixA} for details) with parameters $L_0 = 1.9\:\text{nH}$, $C_0=253\:\text{fF}$, coupling capacitance $(C_v, C_w)=(33, 17)\:\text{fF}$ and mutual inductance $(M_v, M_w)=(-38, -32)\:\text{pH}$ in the trivial phase (the values are interchanged in the case of topological phase). The shaded regions correspond to bandgaps in the dispersion relation of panel \textbf{c}.}\label{fig:Figure1} 
\end{center}
\end{figure*}

The SSH model describing the topological waveguide studied here is illustrated in Fig.~\ref{fig:Figure1}a. Each unit cell of the waveguide consists of two photonic sites, A and B, each containing a resonator with resonant frequency $\omega_0$.  The intra-cell coupling between A and B sites is $J(1 + \delta)$ and the inter-cell coupling between unit cells is $J(1 - \delta)$.  The discrete translational symmetry (lattice constant $d$) of this system allows us to write the Hamiltonian in terms of momentum-space operators, $\hat{H} / \hbar = \sum_{k} (\hat{\mathbf{v}}_k)^\dagger\: \mathbf{h}(k)\:\hat{\mathbf{v}}_k$, where $\hat{\mathbf{v}}_k = (\hat{a}_k,\hat{b}_k)^{T}$ is a vector operator consisting of a pair of A and B sublattice photonic mode operators, and the $k$-dependent kernel of the Hamiltonian is given by,  
\begin{equation}
    {\mathbf{h}}(k) = \begin{pmatrix} \omega_0 & f(k) \\ f^*(k) & \omega_0 \end{pmatrix}.\label{eq:Hamiltonian-SSH}
\end{equation}
Here, $f(k)\equiv - J[(1 + \delta) + (1 - \delta) e^{-i k d} ]$ is the momentum-space coupling between modes on different sublattice, which carries information about the topology of the system. The eigenstates of this Hamiltonian form two symmetric bands centered about the reference frequency $\omega_0$ with dispersion relation

\begin{equation*}
    \omega_{\pm}(k) = \omega_0 \pm J\sqrt{2(1 + \delta^2) + 2(1 - \delta^2) \cos{(k d)}},
\end{equation*}

\noindent where the $+$ ($-$) branch corresponds to the upper (lower) frequency passband. While the band structure is dependent only on the magnitude of $\delta$, and not on whether $\delta>0$ or $\delta<0$, deformation from one case to the other must be accompanied by the closing of the middle bandgap (MBG), defining two topologically distinct phases. For a finite system, it is well known that edge states localized on the boundary of the waveguide at a $\omega=\omega_0$ only appear in the case of $\delta<0$, the so-called \emph{topological} phase~\cite{Asboth:2016, Ozawa:2019}.  The case for which $\delta>0$ is the \emph{trivial} phase with no edge states. It should be noted that for an infinite system, the topological or trivial phase in the SSH model depends on the choice of unit cell, resulting in an ambiguity in defining the bulk properties. Despite this, considering the open boundary of a finite-sized array or a particular section of the bulk, the topological character of the bands can be uniquely defined and can give rise to observable effects. 

We construct a circuit analog of this canonical model using an array of inductor-capacitor (LC) resonators with alternating coupling capacitance and mutual inductance as shown in Fig.~\ref{fig:Figure1}a. The topological phase of the circuit model is determined by the relative size of intra- and inter-cell coupling between neighboring resonators, including both the capacitive and inductive contributions. Strictly speaking, this circuit model breaks chiral symmetry of the original SSH Hamiltonian~\cite{Asboth:2016, Ozawa:2019}, which ensures the band spectrum to be symmetric with respect to $\omega=\omega_0$. Nevertheless, the topological protection of the edge states under perturbation in the intra- and inter-cell coupling strengths remains valid under certain conditions, and the existence of edge states still persists due to the presence of inversion symmetry within the unit cell of the circuit analog, leading to a quantized Zak phase~\cite{Zak:1989}. For detailed analysis of the modeling, symmetry, and robustness of the circuit topological waveguide see Apps.~\ref{sec:AppendixA} and \ref{sec:AppendixB}.

\begin{figure*}[t!]
\begin{center}
\includegraphics[width=1\textwidth]{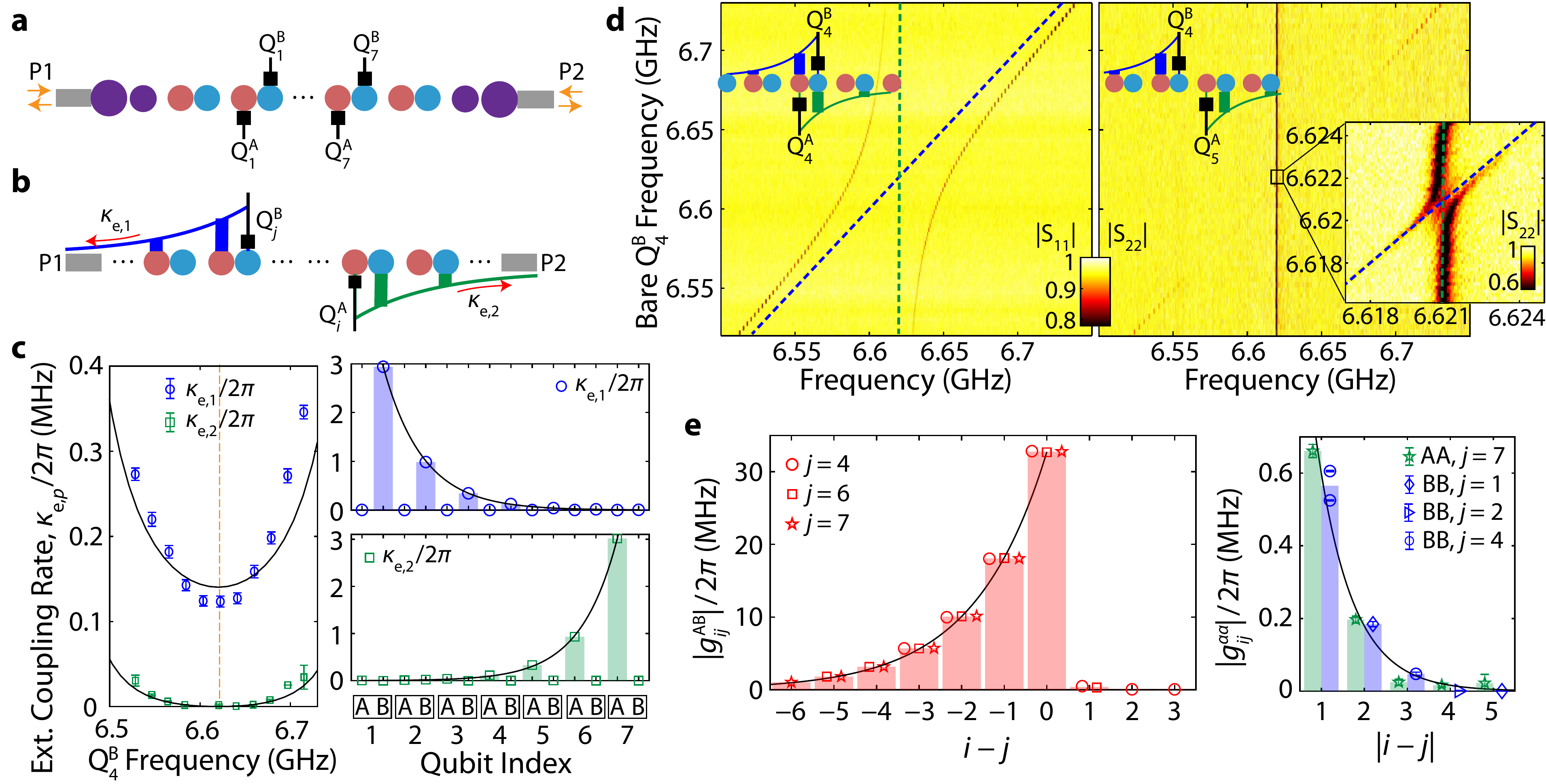}
\caption{\textbf{Directionality of qubit-photon bound states.} \textbf{a}, Schematic of Device I, consisting of 9 unit cells in the trivial phase with qubits (black lines terminated with a square) coupled to every site on the 7 central unit cells. The ends of the array are tapered with additional resonators (purple) with engineered couplings designed to minimize impedance mismatch at upper passband frequencies. \textbf{b}, Theoretical photonic envelope of the directional qubit-photon bound states. At the reference frequency $\omega_0$, the qubit coupled to site A (B) induces a photonic envelope to the right (left), colored in green (blue). The bars along the envelope indicate photon occupation on the corresponding resonator sites. \textbf{c}, Measured coupling rate $\kappa_{\text{e},p}$ to external port numbers, $p=1,2$, of qubit-photon bound states. Left: external coupling rate of qubit Q$_4^\text{B}$ to each port as a function of frequency inside the MBG. Solid black curve is a model fit to the measured external coupling curves.  The frequency point of highest directionality is extracted from the fit curve, and is found to be $\omega_0/2\pi=6.621\:\text{GHz}$ (vertical dashed orange line). Top (Bottom)-right: external coupling rate of all qubits tuned to $\omega=\omega_0$ measured from port P1 (P2). The solid black curves in these plots correspond to exponential fits to the measured external qubit coupling versus qubit index.  \textbf{d}, Two-dimensional color intensity plot of the reflection spectrum under crossing between a pair of qubits with frequency centered around $\omega=\omega_0$. Left: reflection from P1 ($|S_{11}|$) while tuning Q$_4^\text{B}$ across Q$_4^\text{A}$ (fixed). An avoided crossing of $2|g_{44}^\text{AB}|/2\pi = 65.7\:\text{MHz}$ is observed.  Right: reflection from P2 ($|S_{22}|$) while tuning Q$_4^\text{B}$ across Q$_5^\text{A}$ (fixed), indicating the absence of appreciable coupling. Inset to the right shows a zoomed-in region where a small avoided crossing of $2|g_{54}^\text{AB}|/2\pi = 967\:\text{kHz}$ is measured.  The bare qubit frequencies from the fit are shown with dashed lines.  \textbf{e}, Coupling $|g_{ij}^{\alpha\beta}|$ ($\alpha,\beta \in \{\text{A,B}\}$) between various qubit pairs (Q$_{i}^{\alpha}$,Q$_{j}^{\beta}$) at $\omega = \omega_0$, extracted from the crossing experiments similar to panel \textbf{d}. Solid black curves are exponential fits to the measured qubit-qubit coupling rate versus qubit index difference (spatial separation).  Error bars in all figure panels indicate 95\% confidence interval, and are omitted on data points whose marker size is larger than the error itself.}\label{fig:Figure2} 
\end{center}
\end{figure*}

The circuit model is realized using standard fabrication techniques for superconducting metamaterials discussed in Refs.~\cite{Mirhosseini:2018, Ferreira:2020}, where the coupling between sites is controlled by the physical distance between neighboring resonators. Due to the near-field nature, the coupling strength is larger (smaller) for smaller (larger) distance between resonators on a device. An example unit cell of a fabricated device in the topological phase is shown in Fig.~\ref{fig:Figure1}b (the values of intra- and inter-cell distances are interchanged in the trivial phase). We find a good agreement between the measured transmission spectrum and a theoretical curve calculated from a LC lumped-element model of the test structures with 8 unit cells in both trivial and topological configurations (Fig.~\ref{fig:Figure1}c,d). For the topological configuration, the observed peak in the waveguide transmission spectrum at 6.636\:GHz inside the MBG signifies the associated edge state physics in our system.

The non-trivial properties of the topological waveguide can be accessed by coupling quantum emitters to the engineered structure. To this end, we prepare Device I consisting of a topological waveguide in the trivial phase with 9 unit cells, whose boundary is tapered with specially designed resonators before connection to external ports (see Fig.~\ref{fig:Figure2}a). The tapering sections at both ends of the array are designed to reduce the impedance mismatch to the external ports ($Z_0 = 50\:\Omega$) at frequencies in the upper passband (UPB). This is crucial for reducing ripples in the waveguide transmission spectrum in the passbands~\cite{Ferreira:2020}. The device contains 14 frequency-tunable transmon qubits~\cite{Koch:2007} coupled to every site on the 7 unit cells in the middle of the array (labeled Q$_i^\alpha$, where $i=$1-7 and $\alpha$=A,B are the cell and sublattice indices, respectively). Properties of Device I and the tapering section are discussed in further detail in Apps.~\ref{sec:AppendixC} and \ref{sec:AppendixD}, respectively.

For qubits lying within the middle bandgap, the topology of the waveguide manifests itself in the spatial profile of the resulting qubit-photon bound states. When the qubit transition frequency is inside the bandgap, the emission of a propagating photon from the qubit is forbidden due to the absence of photonic modes at the qubit resonant frequency. In this scenario, a stable bound state excitation forms, consisting of a qubit in its excited state and a waveguide photon with exponentially localized photonic envelope~\cite{John:1990,Kurizki:1990}. Generally, bound states with a symmetric photonic envelope emerge due to the inversion symmetry of the photonic bath with respect to the qubit location~\cite{Liu:2016}. In the case of the SSH photonic bath, however, a directional envelope can be realized~\cite{Bello:2019} for a qubit at the centre of the MBG ($\omega_0$), where the presence of a qubit creates a domain wall in the SSH chain and the induced photonic bound state is akin to an edge state (refer to App.~\ref{sec:AppendixE} for a detailed description). For example, in the trivial phase, a qubit coupled to site A (B) acts as the last site of a topological array extended to the right (left) while the subsystem consisting of the remaining sites extended to the left (right) is interpreted as a trivial array. Mimicking the topological edge state, the induced photonic envelope of the bound state faces right (left) with photon occupation only on B (A) sites (Fig.~\ref{fig:Figure2}b), while across the trivial boundary on the left (right) there is no photon occupation. The opposite directional character is expected in the case of the topological phase of the waveguide. The directionality reduces away from the center of the MBG, and is effectively absent inside the upper or lower bandgaps.

We experimentally probe the directionality of qubit-photon bound states by utilizing the coupling of bound states to the external ports in the finite-length waveguide of Device I (see Fig.~\ref{fig:Figure2}c). The external coupling rate $\kappa_{\text{e},p}$ ($p=1,2$) is governed by the overlap of modes in the external port $p$ with the tail of the exponentially attenuated envelope of the bound state, and therefore serves as a useful measure to characterize the localization~\cite{Biondi:2014,Liu:2016, Mirhosseini:2018}. To find the reference frequency $\omega_0$ where the bound state becomes most directional, we measure the external linewidth of the bound state seen from each port as a function of qubit tuning. For Q$_4^\text{B}$, which is located near the center of the array, we find $\kappa_{\text{e},1}$ to be much larger than $\kappa_{\text{e},2}$ at all frequencies inside MBG.  At $\omega_0/2\pi = 6.621\:\text{GHz}$, $\kappa_{\text{e},2}$ completely vanishes, indicating a directionality of the Q$_4^\text{B}$ bound state to the left. Plotting the external coupling at this frequency to both ports against qubit index, we observe a decaying envelope on every other site, signifying the directionality of photonic bound states is correlated with the type of sublattice site a qubit is coupled to. Similar measurements when qubits are tuned to other frequencies near the edge of the MBG, or inside the upper bandgap (UBG), show the loss of directionality away from $\omega=\omega_0$ (App.~\ref{sec:AppendixF}).

A remarkable consequence of the distinctive shape of bound states is direction-dependent photon-mediated interactions between qubits (Fig.~\ref{fig:Figure2}d,e). Due to the site-dependent shapes of qubit-photon bound states, the interaction between qubits becomes substantial only when a qubit on sublattice A is on the left of the other qubit on sublattice B, i.e., $j>i$ for a qubit pair (Q$_{i}^\text{A}$,Q$_{j}^\text{B}$). From the avoided crossing experiments centered at $\omega=\omega_0$, we extract the qubit-qubit coupling as a function of cell displacement $i-j$. An exponential fit of the data gives the localization length of $\xi=1.7$ (in units of lattice constant), close to the estimated value from the circuit model of our system (see App.~\ref{sec:AppendixC}).  While theory predicts the coupling between qubits in the remaining combinations to be zero, we report that coupling of $|g_{ij}^{\text{AA,BB}}|/2\pi \lesssim 0.66\:\text{MHz}$ and $|g_{ij}^\text{AB}|/2\pi \lesssim 0.48\:\text{MHz}$ (for $i>j$) are observed, much smaller than the bound-state-induced coupling, e.g., $|g_{45}^\text{AB}|/2\pi=32.9\:\text{MHz}$. We attribute such spurious couplings to the unintended near-field interaction between qubits. Note that we find consistent coupling strength of qubit pairs dependent only on their relative displacement, not on the actual location in the array, suggesting that physics inside MBG remains intact with the introduced waveguide boundaries. In total, the avoided crossing and external linewidth experiments at $\omega=\omega_0$ provide strong evidence of the shape of qubit-photon bound states, compatible with the theoretical photon occupation illustrated in Fig.~\ref{fig:Figure2}b. 

\begin{figure}[t!]
\begin{center}
\includegraphics[width=0.48\textwidth]{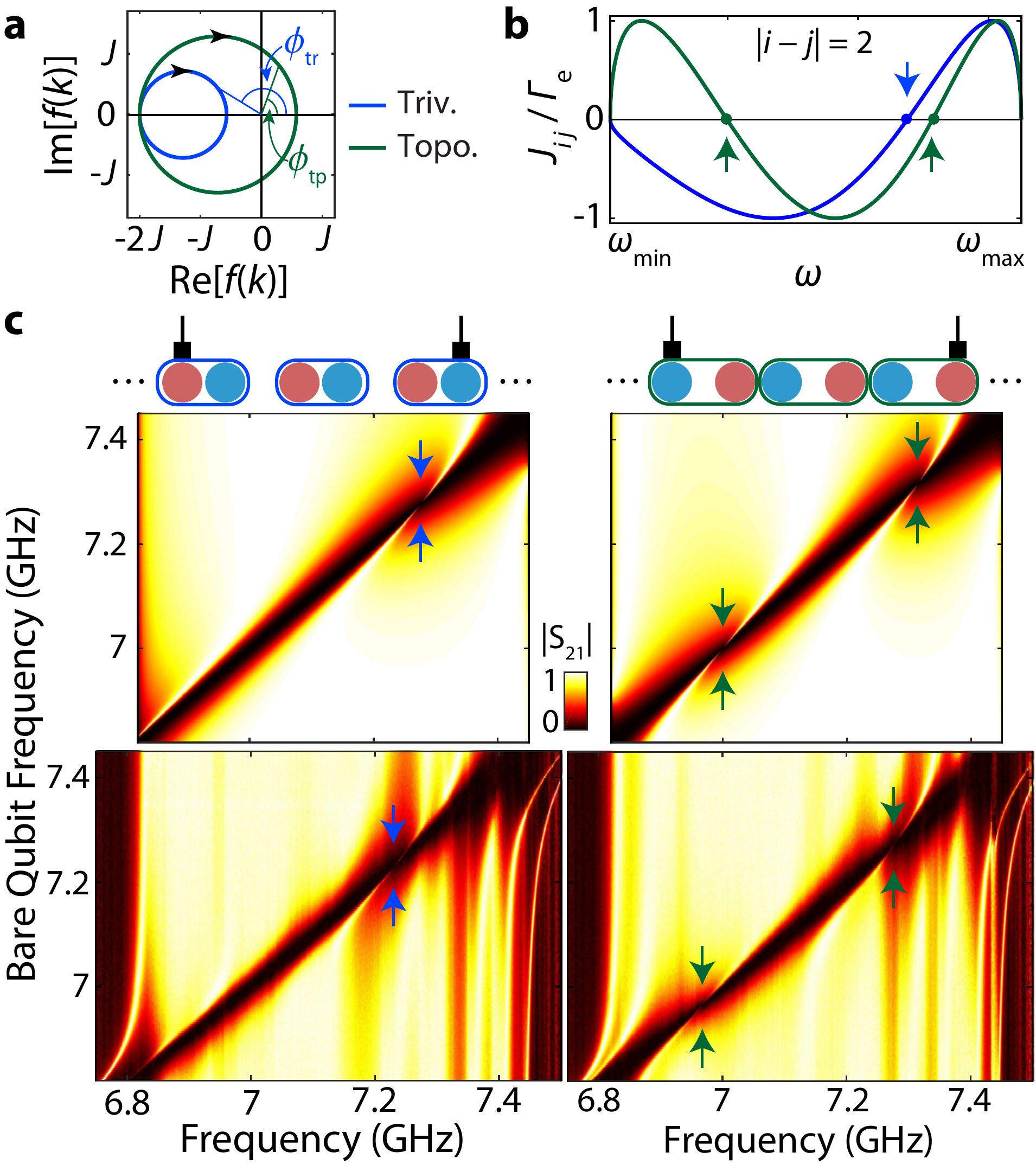}
\caption{\textbf{Probing band topology with qubits.} \textbf{a}, $f(k)$ in the complex plane for $k$ values in the first Brillouin zone. $\phi_\text{tr}$ ($\phi_\text{tp}$) is the phase angle of $f(k)$ for a trivial (topological) section of waveguide, which changes by $0$ ($\pi$) radians as $kd$ transitions from $0$ to $\pi$ (arc in upper plane following black arrowheads). \textbf{b}, Coherent exchange interaction $J_{ij}$ between a pair of coupled qubits as a function of frequency inside the passband, normalized to individual qubit decay rate $\Gamma_\text{e}$ (only $kd \in [0,\pi)$ branch is plotted). Here, one qubit is coupled to the A sublattice of the $i$-th unit cell and the other qubit is coupled to the B sublattice of the $j$-th unit cell, where $|i-j|=2$. Blue (green) curve corresponds to a trivial (topological) intermediate section of waveguide between qubits.  The intercepts at $J_{ij}=0$ (filled circles with arrows) correspond to points where perfect super-radiance occurs. \textbf{c}, Waveguide transmission spectrum $|S_{21}|$ as a qubit pair are resonantly tuned across the UPB of Device I [left: (Q$_{2}^\text{A}$,Q$_{4}^\text{B}$), right: (Q$_{2}^\text{B}$,Q$_{5}^\text{A}$)]. Top: schematic illustrating system configuration during the experiment, with left (right) system corresponding to an interacting qubit pair subtending a three-unit-cell section of waveguide in the trivial (topological) phase. Middle and Bottom two-dimensional color intensity plots of $|S_{21}|$ from theory and experiment, respectively. Swirl patterns (highlighted by arrows) are observed in the vicinity of perfectly super-radiant points, whose number of occurrences differ by one between trivial and topological waveguide sections.}\label{fig:Figure3}
\end{center}
\end{figure}

In the passband regime, i.e., when the qubit frequencies lie within the upper or lower passbands, the topology of the waveguide is imprinted on cooperative interaction between qubits and the single-photon scattering response of the system. The topology of the SSH model can be visualized by plotting the complex-valued $f(k)$ for $k$ values in the first Brillouin zone (Fig.~\ref{fig:Figure3}a). In the topological (trivial) phase, the contour of $f(k)$ encloses (excludes) the origin of the complex plane, resulting in the winding number of 1 (0) and the corresponding Zak phase of $\pi$ (0)~\cite{Zak:1989}. This is consistent with the earlier definition based on the sign of $\delta$. It is known that for a regular waveguide with linear dispersion, the coherent exchange interaction $J_{ij}$ and correlated decay $\Gamma_{ij}$ between qubits at positions $x_i$ and $x_j$ along the waveguide take the forms $J_{ij}\propto \sin{\varphi_{ij}}$ and $\Gamma_{ij}\propto \cos{\varphi_{ij}}$~\cite{Chang:2012, Lalumiere:2013}, where $\varphi_{ij} = k |x_i - x_j|$ is the phase length. In the case of our topological waveguide, considering a pair of qubits coupled to A/B sublattice on $i$/$j$-th unit cell, this argument additionally collects the phase $\phi(k)\equiv \arg {f(k)}$~\cite{Bello:2019}. This is an important difference compared to the regular waveguide case, because the zeros of equation
\begin{equation}
   \varphi_{ij}(k)\equiv kd|i-j| - \phi(k) =0 \mod{\pi}\label{eq:superradiance-eqn}
\end{equation}
determine wavevectors (and corresponding frequencies) where perfect Dicke super-radiance~\cite{Dicke:1954} occurs. Due to the properties of $f(k)$ introduced above, for a fixed cell-distance $\Delta n\equiv|i-j|\ge 1$ between qubits there exists exactly $\Delta n-1$ ($\Delta n$) frequency points inside the passband where perfect super-radiance occurs in the trivial (topological) phase. An example for the $\Delta n=2$ case is shown in Fig.~\ref{fig:Figure3}b. Note that although Eq.~\eqref{eq:superradiance-eqn} is satisfied at the band-edge frequencies $\omega_\text{min}$ and $\omega_\text{max}$ ($kd=\{0,\pi\}$), they are excluded from the above counting due to breakdown of the Born-Markov approximation that is assumed in obtaining the particular form of cooperative interaction in this picture.

To experimentally probe signatures of perfect super-radiance, we tune the frequency of a pair of qubits across the UPB of Device I while keeping the two qubits resonant with each other. We measure the waveguide transmission spectrum $S_{21}$ during this tuning, keeping track of the lineshape of the two-qubit resonance as $J_{ij}$ and $\Gamma_{ij}$ varies over the tuning. Drastic changes in the waveguide transmission spectrum occur whenever the two-qubit resonance passes through the perfectly super-radiant points, resulting in a swirl pattern in $|S_{21}|$. Such patterns arise from the disappearance of the peak in transmission associated with interference between photons scattered by imperfect super- and sub-radiant states, resembling the electromagnetically-induced transparency in a V-type atomic level structure~\cite{Witthaut:2010}. As an example, we discuss the cases with qubit pairs (Q$_2^\text{A}$,Q$_4^\text{B}$) and (Q$_2^\text{B}$,Q$_5^\text{A}$), which are shown in Fig.~\ref{fig:Figure3}c.  Each qubit pair configuration encloses a three-unit-cell section of the waveguide; however for the (Q$_2^\text{A}$,Q$_4^\text{B}$) pair the waveguide section is in the trivial phase, whereas for (Q$_2^\text{A}$,Q$_4^\text{B}$) the waveguide section is in the topological phase. Both theory and measurement indicate that the qubit pair (Q$_2^\text{A}$,Q$_4^\text{B}$) has exactly one perfectly super-radiant frequency point in the UPB. For the other qubit pair (Q$_2^\text{B}$,Q$_5^\text{A}$), with waveguide section in the topological phase, two such points occur (corresponding to $\Delta n=2$). This observation highlights the fact that while the topological phase of the bulk in the SSH model is ambiguous, a finite section of the array can still be interpreted to have a definite topological phase. Apart from the unintended ripples near the band-edges, the observed lineshapes are in good qualitative agreement with the theoretical expectation in Ref.~\cite{Bello:2019}. Detailed description of the swirl pattern and similar measurement results for other qubit combinations with varying $\Delta n$ are reported in App.~\ref{sec:AppendixG}.

\begin{figure}[t!]
\begin{center}
\includegraphics[width=0.48\textwidth]{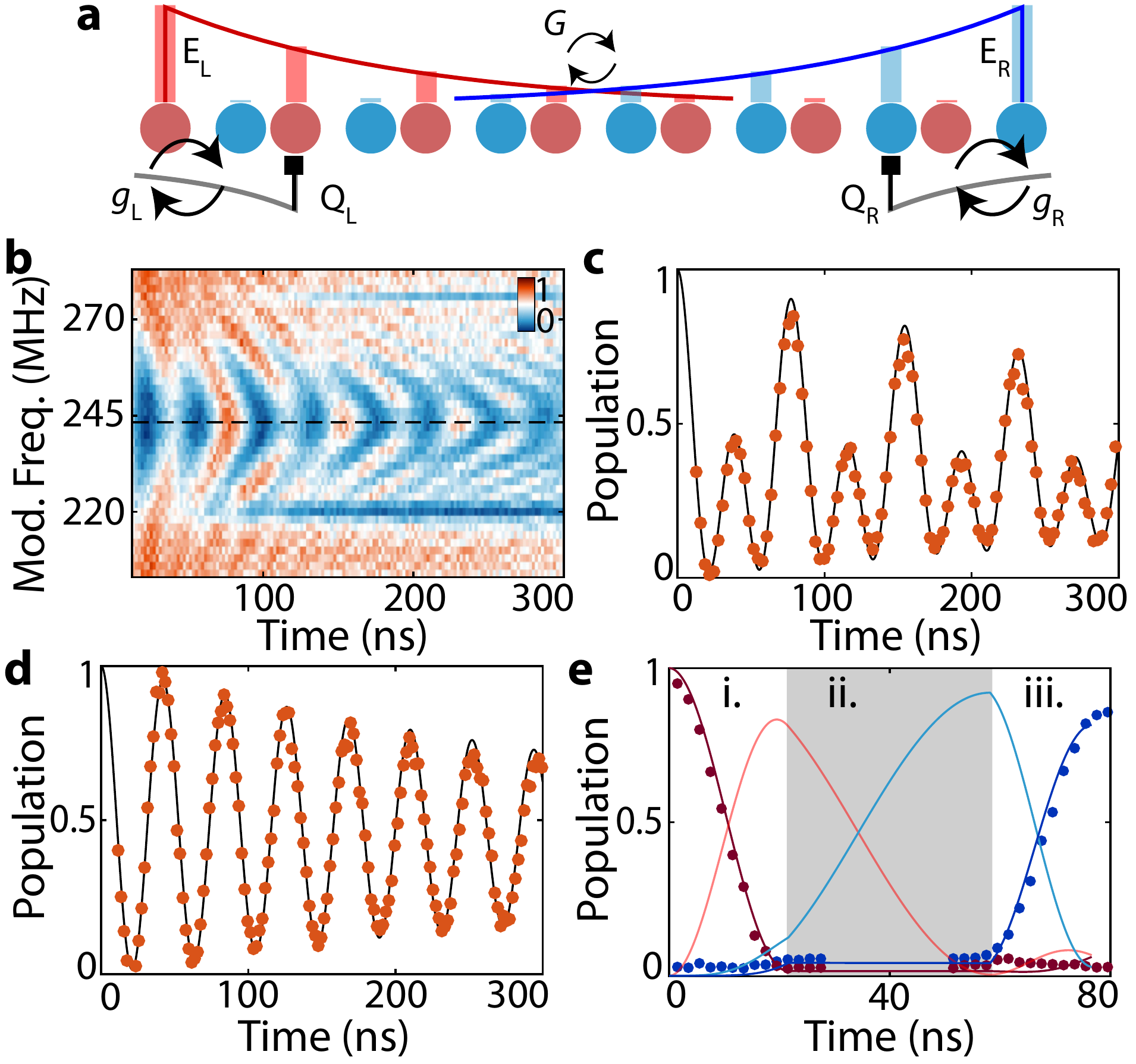}
\caption{\textbf{Qubit interaction with topological edge modes.} \textbf{a}, Schematic of Device II, consisting of 7 unit cells in the topological phase with qubits \QL{}$=\text{Q}_{i}^{\alpha}$ and \QR{}$=\text{Q}_{j}^{\beta}$ coupled at sites $(i,\alpha)=(2,\text{A})$ and $(j,\beta)=(6,\text{B})$, respectively. \EL{} and \ER{} are the left-localized and right-localized edge modes which interact with each other at rate $G$ due to their overlap in the center of the finite waveguide. \textbf{b}, Chevron-shaped oscillation of \QL{} population arising from interaction with edge modes under variable frequency and duration of modulation pulse. The oscillation is nearly symmetric with respect to optimal modulation frequency $242.5$\:MHz, apart from additional features at ($219$, $275$)\:MHz due to spurious interaction of unused sidebands with modes inside the passband. \textbf{c}, Line-cut of panel b (indicated with a dashed line) at the optimal modulation frequency. A population oscillation involving two harmonics is observed due to coupling of \EL{} to \ER{}. \textbf{d}, Vacuum Rabi oscillations between \QL{} and \EL{} when \QR{} is parked at the resonant frequency of edge modes to shift the frequency of \ER{}, during the process in panel \textbf{c}. In panels \textbf{c} and \textbf{d}, the filled orange circles (black solid lines) are the data from experiment (theory). \textbf{e}, Population transfer from \QL{} to \QR{} composed of three consecutive swap transfers \QL$\rightarrow$\EL$\rightarrow$\ER$\rightarrow$\QR. The population of \QL{} (\QR{}) during the process is colored dark red (dark blue), with filled circles and solid lines showing the measured data and fit from theory, respectively. The light red (light blue) curve indicates the expected population in \EL{} (\ER{}) mode, calculated from theory.}\label{fig:Figure4}
\end{center}
\end{figure}

Finally, to explore the physics associated with topological edge modes, we fabricated a second device, Device II, which realizes a closed quantum system with 7 unit cells in the topological phase (Fig.~\ref{fig:Figure4}a). We denote the photonic sites in the array by ($i$,$\alpha$), where $i=$1-7 is the cell index and $\alpha=$A,B is the sublattice index. Due to reflection at the boundary, the passbands on this device appear as sets of discrete resonances. The system supports topological edge modes localized near the sites (1,A) and (7,B) at the boundary, labeled \EL\ and \ER. The edge modes are spatially distributed with exponentially attenuated tails directed toward the bulk. In a finite system, the non-vanishing overlap between the envelopes of edge states generates a coupling which depends on the localization length $\xi$ and the system size $L$ as $G\sim e^{-L/\xi}$. In Device II, two qubits denoted \QL{} and \QR{} are coupled to the topological waveguide at sites (2,A) and (6,B), respectively. Each qubit has a local drive line and a flux-bias line, which are connected to room-temperature electronics for control. The qubits are dispersively coupled to readout resonators, which are loaded to a coplanar waveguide for time-domain measurement. The edge mode \EL{} (\ER{}) has photon occupation on sublattice A (B), inducing interaction \gL{} (\gR{}) with \QL{} (\QR{}). Due to the directional properties discussed earlier, bound states arising from \QL{} and \QR{} have photonic envelopes facing away from each other inside the MBG, and hence have no direct coupling to each other. For additional details on Device II and qubit control, refer to App.~\ref{sec:AppendixH}.

We probe the topological edge modes by utilizing the interaction with the qubits. While parking \QL{} at frequency $f_\text{q}=6.835\:\text{GHz}$ inside MBG, we initialize the qubit into its excited state by applying a microwave $\pi$-pulse to the local drive line. Then, the frequency of the qubit is parametrically modulated~\cite{Naik:2017} such that the first-order sideband of the qubit transition frequency is nearly resonant with \EL{}. After a variable duration of the frequency modulation pulse, the state of the qubit is read out. From this measurement, we find a chevron-shaped oscillation of the qubit population in time centered at modulation frequency $242.5\:\text{MHz}$ (Fig.~\ref{fig:Figure4}b). We find the population oscillation at this modulation frequency to contain two harmonic components as shown in Fig.~\ref{fig:Figure4}c, a general feature of a system consisting of three states with two exchange-type interactions $g_1$ and $g_2$. In such cases, three single-excitation eigenstates exist at $0$, $\pm {g}$ with respect to the bare resonant frequency of the emitters (${g}\equiv \sqrt{g_1^2+g_2^2}$), and since the only possible spacing between the eigenstates in this case is ${g}$ and $2{g}$, the dynamics of the qubit population exhibits two frequency components with a ratio of two. From fitting the \QL{} population oscillation data in Fig.~\ref{fig:Figure4}c, the coupling between \EL{} and \ER{} is extracted to be $G/2\pi=5.05\:\text{MHz}$. Parking \QR{} at the bare resonant frequency $\omega_\text{E}/2\pi=6.601\:\text{GHz}$ of the edge modes, \ER{} strongly hybridizes with \QR{} and is spectrally distributed at $\pm g_\text{R}$ with respect to the original frequency ($g_\text{R}/2\pi=57.3\:\text{MHz}$). As this splitting is much larger than the coupling of \ER{} to \EL{}, the interaction channel \EL$\leftrightarrow$\ER{} is effectively suppressed and the vacuum Rabi oscillation only involving \QL{} and \EL{} is recovered (Fig.~\ref{fig:Figure4}d) by applying the above-mentioned pulse sequence on \QL{}. A similar result was achieved by applying a simultaneous modulation pulse on \QR{} to put its first-order sideband near-resonance with the bare edge modes (instead of parking it near resonance), which we call the \emph{double-modulation} scheme. From the vacuum Rabi oscillation \QL$\leftrightarrow$\EL{} (\QR$\leftrightarrow$\ER) using the double-modulation scheme, we find the effective qubit-edge mode coupling to be $\tilde{g}_\text{L}/2\pi = 23.8\:\text{MHz}$ ($\tilde{g}_\text{R}/2\pi = 22.5\:\text{MHz}$).

The half-period of vacuum Rabi oscillation corresponds to an iSWAP gate between \QL{} and \EL{} (or \QR{} and \ER{}), which enables control over the edge modes with single-photon precision. As a demonstration of this tool, we perform remote population transfer between \QL{} and \QR through the non-local coupling of topological edge modes \EL\ and \ER. The qubit \QL{} (\QR{}) is parked at frequency 6.829\:GHz (6.835\:GHz) and prepared in its excited (ground) state. The transfer protocol, consisting of three steps, is implemented as follows: i) an iSWAP gate between \QL{} and \EL{} is applied by utilizing the vacuum Rabi oscillation during the double-modulation scheme mentioned above, ii) the frequency modulation is turned off and population is exchanged from \EL{} to \ER{} using the interaction $G$, iii) another iSWAP gate between \QR{} and \ER{} is applied to map the population from \ER{} to \QR. The population of both qubits at any time within the transfer process is measured using multiplexed readout~\cite{Chen:2012} (Fig.~\ref{fig:Figure4}e). We find the final population in \QR{} after the transfer process to be 87\:\%. Numerical simulations suggest that (App.~\ref{sec:AppendixH}) the infidelity in preparing the initial excited state accounts for 1.6\:\% of the population decrease, the leakage to the unintended edge mode due to ever-present interaction $G$ contributes 4.9\:\%, and the remaining 6.5\:\% is ascribed to the short coherence time of qubits away from the flux-insensitive point [$T_2^*=344\:(539)$\:ns for \QL{} (\QR) at working point].

We expect that a moderate improvement on the demonstrated population transfer protocol could be achieved by careful enhancement of the excited state preparation and the iSWAP gates, i.e. optimizing the shapes of the control pulses~\cite{Khaneja:2005, Motzoi:2009, Didier:2018, Hong:2020}. The coherence-limited infidelity can be mitigated by utilizing a less flux-sensitive qubit design~\cite{Sete:2017, Hutchings:2017} or by reducing the generic noise level of the experimental setup~\cite{Ott:2009}. Further, incorporating tunable couplers~\cite{Chen:2014} into the existing metamaterial architecture to control the localization length of edge states \emph{in situ} will fully address the population leakage into unintended interaction channels, and more importantly, enable robust quantum state transfer over long distances~\cite{Lang:2017}. Together with many-body protection to enhance the robustness of topological states~\cite{deLeseleuc:2019}, building blocks of quantum communication~\cite{Kimble:2008} under topological protection are also conceivable. 

Looking forward, we envision several research directions to be explored beyond the work presented here. First, the topology-dependent photon scattering in photonic bands that is imprinted in the cooperative interaction of qubits can lead to new ways of measuring topological invariants in photonic systems~\cite{Tran:2017}. The directional and long-range photon-mediated interactions between qubits demonstrated in our work also opens avenues to synthesize  non-trivial quantum many-body states of qubits, such as the double N\'eel state~\cite{Bello:2019}. Even without technical advances in fabrication~\cite{Dunsworth:2018,Rosenberg:2017,Foxen:2017}, a natural scale-up of the current system will allow for the construction of moderate to large-scale quantum many-body systems. Specifically, due to the on-chip wiring efficiency of a linear waveguide QED architecture, with realistic refinements involving placement of local control lines on qubits and compact readout resonators coupled to the tapered passband (intrinsically acting as Purcell filters~\cite{Jeffrey:2014}), we expect that a fully controlled quantum many-body system consisting of 100 qubits is realizable in the near future. In such systems, protocols for preparing and stabilizing~\cite{deLeseleuc:2019, Ma:2017, Ma:2019}  quantum many-body states could be utilized and tested. Additionally, the flexibility of superconducting metamaterial architectures~\cite{Mirhosseini:2018, Ferreira:2020} can be further exploited to realize other novel types of topological photonic baths~\cite{Lodahl:2017, Bello:2019, Garcia-Elcano:2019}. While the present work was limited to a one-dimensional system, the state-of-the-art technologies in superconducting quantum circuits~\cite{Krantz:2019} utilizing flip-chip methods~\cite{Rosenberg:2017,Foxen:2017} will enable integration of qubits into two-dimensional metamaterial surfaces. It also remains to be explored whether topological models with broken time-reversal symmetry, an actively pursued approach in systems consisting of arrays of three-dimensional microwave cavities~\cite{Anderson:2016, Owens:2018}, could be realized in compact chip-based architectures. Altogether, our work sheds light on opportunities in superconducting circuits to explore quantum many-body physics originating from novel types of photon-mediated interactions in topological waveguide QED, and paves the way for creating synthetic quantum matter and performing quantum simulation~\cite{Blatt:2012,Gross:2017,Lewis-Swan:2019,Georgescu:2014,Carusotto:2020}.

\begin{acknowledgments}
The authors thank Xie Chen and Hans Peter B\"uchler for helpful discussions. We also appreciate MIT Lincoln Laboratories for the provision of a traveling-wave parametric amplifier used for both spectroscopic and time-domain measurements in this work, and Jen-Hao Yeh and B.~S.~Palmer for the cryogenic attenuators for reducing thermal noise in the metamaterial waveguide. This work was supported by the AFOSR MURI Quantum Photonic Matter (grant FA9550-16-1-0323), the DOE-BES Quantum Information Science Program (grant DE-SC0020152), the AWS Center for Quantum Computing, the Institute for Quantum Information and Matter, an NSF Physics Frontiers Center (grant PHY-1125565) with support of the Gordon and Betty Moore Foundation, and the Kavli Nanoscience Institute at Caltech. V.F.~gratefully acknowledges support from NSF GFRP Fellowship. A.S.~is supported by IQIM Postdoctoral Fellowship. A.G.-T.~acknowledges funding from project PGC2018-094792-B-I00  (MCIU/AEI/FEDER, UE), CSIC Research Platform PTI-001, and CAM/FEDER Project No.~S2018/TCS-4342~(QUITEMAD-CM).
  
\end{acknowledgments}

\appendix


\section{Modeling of the topological waveguide}
\label{sec:AppendixA}
\begin{figure}[b]
\centering
\includegraphics[width=0.5\textwidth]{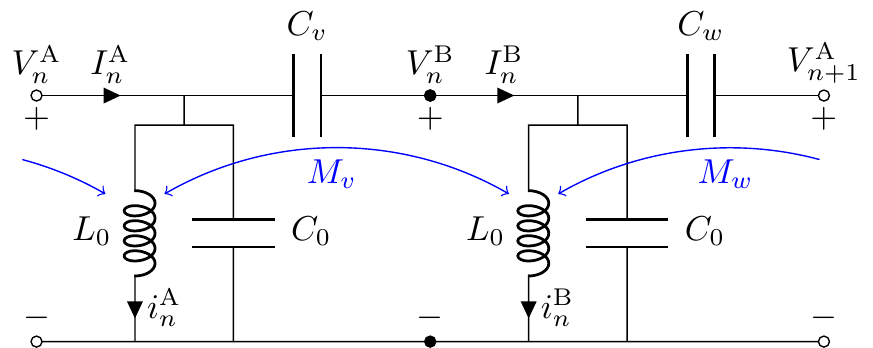}
  \caption{\textbf{Circuit model of the topological waveguide unit cell.} LC resonators of inductance $L_0$ and capacitance $C_0$ are coupled with alternating coupling capacitance $C_v$, $C_w$ and mutual inductance $M_v$, $M_w$. The voltage and current at each resonator node A (B) are denoted as $V_n^\text{A},I_n^\text{A}$ ($V_n^\text{B},I_n^\text{B}$).}
  \label{fig:FigureS1}
\end{figure}
In this section we provide a theoretical description of the topological waveguide discussed in the main text, an analog to the Su-Schrieffer-Heeger model~\cite{Su:1979}.  An approximate form of the physically realized waveguide is given by an array of coupled LC resonators, a unit cell of which is illustrated in Fig.~\ref{fig:FigureS1}. Each unit cell of the topological waveguide has two sites A and B whose intra- and inter-cell coupling capacitance (mutual inductance) are given by $C_v$ ($M_v$) and $C_w$ ($M_w$). We denote the flux variable of each node as $\Phi_n^{\alpha}(t)\equiv \int_{-\infty}^{t} \ud t' \: V_n^\alpha(t')$ and the current going through each inductor as $i_n^{\alpha}$ ($\alpha=\{\text{A},\text{B}\}$). The Lagrangian in position space reads
\begin{widetext}
\begin{align}
    \mathcal{L} = & \sum_n\Bigg\{\frac{C_v}{2}\left(\dot{\Phi}_n^\text{B} - \dot{\Phi}_n^\text{A}\right)^2 + \frac{C_w}{2}\left(\dot{\Phi}_{n+1}^\text{A} - \dot{\Phi}_n^\text{B}\right)^2 + \frac{C_0}{2}\left[\left(\dot{\Phi}_n^\text{A}\right)^2+\left(\dot{\Phi}_n^\text{B}\right)^2\right] \nonumber\\  &\quad\quad\quad -\frac{L_0}{2}\left[\left(i_n^\text{A}\right)^2 + \left(i_n^\text{B}\right)^2\right] - M_v i_n^\text{A} i_n^\text{B} - M_w i_n^\text{B} i_{n+1}^\text{A}\Bigg\}.
    \label{eq:Lagn_new}
\end{align}
The node flux variables are written in terms of current through the inductors as
\begin{align}
    \Phi_n^\text{A} = L_0 i_n^\text{A} + M_v i_n^\text{B} + M_w i_{n-1}^\text{B},\quad
    \Phi_n^\text{B} = L_0 i_n^\text{B} + M_v i_n^\text{A} + M_w i_{n+1}^\text{A}.\label{eq:Phi2i}
\end{align}
Considering the discrete translational symmetry in our system, we can rewrite the variables in terms of Fourier components as
\begin{equation}
    \Phi_n^\alpha = \frac{1}{\sqrt{N}}\sum_{k}e^{inkd}\Phi_k^\alpha, \quad    i_n^\alpha = \frac{1}{\sqrt{N}}\sum_{k}e^{inkd}i_k^\alpha, \label{eq:fourier-transform}
\end{equation}
where $\alpha = \text{A},\text{B}$, $N$ is the number of unit cells, and $k=\frac{2\pi m}{Nd}$ ($m=-N/2, \cdots,N/2-1$) are points in the first Brillouin zone. Equation \eqref{eq:Phi2i} is written as
\begin{equation*}
    \sum_{k'}e^{ink'd}\Phi_{k'}^\text{A}= \sum_{k'}e^{ink'd}\left(L_0 i_{k'}^\text{A} + M_v i_{k'}^\text{B} + e^{-ik'd} M_w i_{k'}^\text{B}\right)
\end{equation*}
under this transform. Multiplying the above equation with $e^{-inkd}$ and summing over all $n$, we get a linear relation between $\Phi_k^\alpha$ and $i_k^\alpha$:
\begin{equation*}
    \begin{pmatrix}
    \Phi_k^\text{A} \\
    \Phi_k^\text{B}
    \end{pmatrix} = 
    \begin{pmatrix}
    L_0 & M_v + M_w e^{-ikd} \\
    M_v + M_w e^{ikd} & L_0
    \end{pmatrix}
    \begin{pmatrix}
    i_k^\text{A} \\
    i_k^\text{B}
    \end{pmatrix}.
\end{equation*}
By calculating the inverse of this relation, the Lagrangian of the system \eqref{eq:Lagn_new} can be rewritten in $k$-space as
\begin{align}
    \mathcal{L} &= \sum_{k} \bigg[ \frac{C_0 + C_v + C_w}{2}\left(\dot{\Phi}_{-k}^\text{A} \dot{\Phi}_{k}^\text{A} + \dot{\Phi}_{-k}^\text{B} \dot{\Phi}_{k}^\text{B}\right) 
    - C_g(k)\:\dot{\Phi}_{-k}^\text{A}\dot{\Phi}_{k}^\text{B} - \frac{L_0}{2}\left(i_{-k}^\text{A}i_{k}^\text{A} + i_{-k}^\text{B}i_{k}^\text{B}\right)-M_g(k)\:i_{-k}^\text{A}i_{k}^\text{B} \bigg]\nonumber\\
    &=\sum_{k}\bigg[\frac{C_0 + C_v + C_w}{2}\left(\dot{\Phi}_{-k}^\text{A} \dot{\Phi}_{k}^\text{A} + \dot{\Phi}_{-k}^\text{B} \dot{\Phi}_{k}^\text{B}\right) 
    - C_g(k)\:\dot{\Phi}_{-k}^\text{A}\dot{\Phi}_{k}^\text{B} - \frac{\frac{L_0}{2}\left(\Phi_{-k}^\text{A}\Phi_{k}^\text{A}+\Phi_{-k}^\text{B}\Phi_{k}^\text{B}\right) -  M_g(k)\Phi_{-k}^\text{A}\Phi_{k}^\text{B} }{L_0^2 - M_g(-k)M_g(k)}\bigg]\label{eq:Lagrangian-kspace}
\end{align}
where $C_g(k) \equiv C_v +C_w e^{-ikd}$ and $M_g(k)\equiv M_v +M_w e^{-ikd}$. The node charge variables $Q_{k}^\alpha\equiv\partial \mathcal{L}/\partial \dot{\Phi}_k^\alpha$ canonically conjugate to node flux $\Phi_k^\alpha$ are 
\begin{align*}
    \begin{pmatrix}
    Q_{k}^\text{A} \\ Q_{k}^\text{B}
    \end{pmatrix}
    =
    \begin{pmatrix}
    C_0 + C_v + C_w & -C_g(-k) \\
    -C_g(k) & C_0 + C_v + C_w
    \end{pmatrix}
    \begin{pmatrix}
    \dot{\Phi}_{-k}^\text{A} \\ \dot{\Phi}_{-k}^\text{B}
    \end{pmatrix}.
\end{align*}
Note that due to the Fourier transform implemented on flux variables, the canonical charge in momentum space is related to that in real space by
\begin{equation*}
    Q_n^\alpha = \frac{\partial \mathcal{L}}{\partial \dot{\Phi}_n^\alpha} = \sum_{k}\frac{\partial \mathcal{L}}{\partial \dot{\Phi}_k^\alpha} \frac{ \partial \dot{\Phi}_k^\alpha}{ \partial \dot{\Phi}_n^\alpha} = \frac{1}{\sqrt{N}}\sum_{k}e^{-inkd}{Q}_k^\alpha,
\end{equation*}
which is in the opposite sense of regular Fourier transform in Eq.~\eqref{eq:fourier-transform}. Also, due to the Fourier-transform properties, the constraint that $\Phi_n^\alpha$ and $Q_n^\alpha$ are real reduces to $(\Phi_k^\alpha)^*=\Phi_{-k}^\alpha$ and $(Q_k^\alpha)^*=Q_{-k}^\alpha$.
Applying the Legendre transformation $H=\sum_{k,\alpha}Q_k^\alpha\dot{\Phi}_{k}^\alpha -\mathcal{L}$, the Hamiltonian takes the form
\begin{align*}
    H &=
\sum_k\bigg[\frac{C_\Sigma (Q_{-k}^\text{A}Q_{k}^\text{A} + Q_{-k}^\text{B}Q_{k}^\text{B}) + C_g(-k)Q_{-k}^\text{A}Q_{k}^\text{B} + C_g(k)Q_{-k}^\text{B}Q_{k}^\text{A}}{2C_d^2(k)}
       \nonumber \\ &\quad\quad\quad\quad + \frac{L_0(\Phi_{-k}^\text{A}\Phi_{k}^\text{A} + \Phi_{-k}^\text{B}\Phi_{k}^\text{B})-M_g(k)\Phi_{-k}^\text{A}\Phi_k^\text{B} - M_g(-k)\Phi_{-k}^\text{B}\Phi_k^\text{A}}{2L_d^2(k)}\bigg],
\end{align*}
where
\begin{align*}
    C_\Sigma \equiv C_0 + C_v + C_w,\quad
    C_d^2(k) \equiv C_\Sigma^2 - C_g(-k)C_g(k),\quad
    L_d^2(k) \equiv L_0^2 - M_g(-k) M_g(k).
\end{align*}
Note that $C_d^2(k)$ and $L_d^2(k)$ are real and even function in $k$. We impose the canonical commutation relation between real-space conjugate variables $[\hat{\Phi}_{n}^\alpha, \hat{Q}_{n'}^\beta]=i\hbar \delta_{\alpha,\beta}\delta_{n,n'}$ to promote the flux and charge variables to quantum operators. This reduces to $[\hat{\Phi}_{k}^\alpha, \hat{Q}_{k'}^\beta]=i\hbar \delta_{\alpha,\beta}\delta_{k,k'}$ in the momentum space [Note that due to the Fourier transform, $(\hat{\Phi}_k^\alpha)^\dagger = \hat{\Phi}_{-k}^\alpha$ and $(\hat{Q}_k^\alpha)^\dagger = \hat{Q}_{-k}^\alpha$, meaning flux and charge operators in momentum space are \emph{non-Hermitian} since the Hermitian conjugate flips the sign of $k$]. The Hamiltonian can be written as a sum $\hat{H}=\hat{H}_0 + \hat{V}$, where the ``uncoupled'' part $\hat{H}_0$ and coupling terms $\hat{V}$ are written as
\begin{align}
    \hat{H}_0 = \sum_{k,\alpha} \left[\frac{\hat{Q}_{-k}^{\alpha} \hat{Q}_{k}^{\alpha}}{2C^\text{eff}_0(k)}  + \frac{\hat{\Phi}_{-k}^\alpha \hat{\Phi}_{k}^\alpha}{2L^\text{eff}_0(k)}\right],\quad \hat{V} = \sum_{k}\left[\frac{\hat{Q}_{-k}^\text{A}\hat{Q}_{k}^\text{B}}{2C_g^\text{eff}(k)} + \frac{\hat{\Phi}_{-k}^\text{A}\hat{\Phi}_{k}^\text{B}}{2L_g^\text{eff}(k)} + \mathrm{H.c.}\right],\label{eq:Hamiltonian-kspace-QPhi}
\end{align}
with the effective self-capacitance $C_0^\text{eff}(k)$, self-inductance $L_0^\text{eff}(k)$, coupling capacitance $C_g^\text{eff}(k)$, and coupling inductance $L_g^\text{eff}(k)$ given by
\begin{align}
    C_0^\text{eff}(k) = \frac{C_d^2(k)}{C_\Sigma}, \quad L_0^\text{eff}(k) = \frac{L_d^2(k)}{L_0},\quad C_g^\text{eff}(k) = \frac{C_d^2(k)}{C_g(-k)},\quad L_g^\text{eff}(k) = -\frac{L_d^2(k)}{M_g(k)}.\label{eq:effective-cap-ind}
\end{align}
The diagonal part $\hat{H}_0$ of the Hamiltonian can be written in a second-quantized form by introducing annihilation operators $\hat{a}_k$ and $\hat{b}_k$, which are operators of the Bloch waves on A and B sublattice, respectively:
\begin{align*}
    \hat{a}_k \equiv \frac{1}{\sqrt{2\hbar}} \left[\frac{\hat{\Phi}_k^\text{A}}{\sqrt{Z_0^\text{eff}(k)} } + i\sqrt{Z_0^\text{eff}(k)}\hat{Q}_{-k}^\text{A} \right],\quad
    \hat{b}_k \equiv \frac{1}{\sqrt{2\hbar}} \left[\frac{\hat{\Phi}_k^\text{B}}{\sqrt{Z_0^\text{eff}(k)} } + i\sqrt{Z_0^\text{eff}(k)} \hat{Q}_{-k}^\text{B} \right].
\end{align*}
Here, $Z_0^\text{eff}(k) \equiv \sqrt{L_0^\text{eff}(k)/C_0^\text{eff}(k)}$ is the effective impedance of the oscillator at wavevector $k$. Unlike the Fourier transform notation, for bosonic modes $\hat{a}_k$ and $\hat{b}_k$, we use the notation $(\hat{a}_k)^\dagger \equiv \hat{a}_k^\dagger$ and $(\hat{b}_k)^\dagger \equiv \hat{b}_k^\dagger$. Under this definition, the commutation relation is rewritten as $[\hat{a}_k,\hat{a}^\dagger_{k'}] = [\hat{b}_k,\hat{b}_{k'}^\dagger]= \delta_{k,k'}$. Note that the flux and charge operators are written in terms of mode operators as
\begin{align*}
    \hat{\Phi}_k^\text{A} &= \sqrt{\frac{\hbar Z_0^\text{eff}(k)}{2}}\left(\hat{a}_k + \hat{a}^\dagger_{-k}\right),\quad
    \hat{Q}_k^\text{A} = \frac{1}{i}\sqrt{\frac{\hbar }{2 Z_0^\text{eff}(k)}}\left(\hat{a}_{-k} - \hat{a}^\dagger_{k}\right),\\
     \hat{\Phi}_k^\text{B} &= \sqrt{\frac{\hbar Z_0^\text{eff}(k)}{2}}\left(\hat{b}_k + \hat{b}^\dagger_{-k}\right),\quad 
     \hat{Q}_k^\text{B} = \frac{1}{i}\sqrt{\frac{\hbar }{2 Z_0^\text{eff}(k)}}\left(\hat{b}_{-k} - \hat{b}^\dagger_{k}\right).
\end{align*}
The uncoupled Hamiltonian is written as
\begin{align}
\hat{H}_0 = \sum_{k}\frac{\hbar\omega_{0}(k)}{2}\left( \hat{a}_k^\dagger \hat{a}_k + \hat{a}_{-k} \hat{a}_{-k}^\dagger + \hat{b}_k^\dagger \hat{b}_k + \hat{b}_{-k} \hat{b}_{-k}^\dagger\right), \label{eq:H0}
\end{align}
where the ``uncoupled'' oscillator frequency is given by $\omega_0(k) \equiv [L_0^\text{eff}(k)C_0^\text{eff}(k)]^{-1/2}$, which ranges between values
$$
\omega_0(k=0)=\sqrt{\frac{L_0 C_\Sigma}{[L_0^2 - (M_v + M_w)^2][C_\Sigma^2 - (C_v + C_w)^2]}}
,\quad
\omega_0 \left(k=\frac{\pi}{d}\right)=\sqrt{\frac{L_0 C_\Sigma}{(L_0^2 - |M_v - M_w|^2)(C_\Sigma^2 - |C_v - C_w|^2)}}.
$$
The coupling Hamiltonian $\hat{V}$ is rewritten as
\begin{align}
    \hat{V} = -\sum_{k}\left[\frac{\hbar g_C(k)}{2}\left(\hat{a}_{-k}\hat{b}_k - \hat{a}_{-k}\hat{b}_{-k}^\dagger - \hat{a}_{k}^\dagger \hat{b}_k + \hat{a}_{k}^\dagger \hat{b}_{-k}^\dagger\right) 
    + \frac{\hbar g_L(k)}{2}\left(\hat{a}_{-k}\hat{b}_k + \hat{a}_{-k}\hat{b}_{-k}^\dagger + \hat{a}_{k}^\dagger\hat{b}_k + \hat{a}_{k}^\dagger\hat{b}_{-k}^\dagger\right) +\mathrm{H.c.}\right],\label{eq:V}
\end{align}
\end{widetext}
where the capacitive coupling $g_C(k)$ and inductive coupling $g_L(k)$ are simply written as 
\begin{equation}
   g_C(k) = \frac{\omega_0(k) C_g(k)}{2C_\Sigma},\quad g_L(k)=\frac{\omega_0(k) M_g(k)}{2L_0},\label{eq:coupling-terms}
\end{equation} respectively. Note that $g_C^*(k)=g_C(-k)$ and $g_L^*(k)=g_L(-k)$. In the following, we discuss the diagonalization of this Hamiltonian to explain the dispersion relation and band topology.

\subsection{Band structure within the rotating-wave approximation}
We first consider the band structure of the system within the rotating-wave approximation (RWA), where we discard the counter-rotating terms $\hat{a}\hat{b}$ and $\hat{a}^\dagger\hat{b}^\dagger$ in the Hamiltonian. This assumption is known to be valid when the strength of the couplings $|g_L(k)|$, $|g_C(k)|$ are small compared to the uncoupled oscillator frequency $\omega_0(k)$. Under this approximation, the Hamiltonian in Eqs.~\eqref{eq:H0}-\eqref{eq:V} reduces to a simple form $\hat{H} = \hbar\sum_{k}(\hat{\mathbf{v}}_k)^\dagger \mathbf{h}(k) \hat{\mathbf{v}}_k$, where the single-particle kernel of the Hamiltonian is,
\begin{equation}
    {\mathbf{h}}(k) =
    \begin{pmatrix}
    \omega_0(k) &  f(k)\\
     f^*(k) & \omega_0(k)
    \end{pmatrix}.\label{eq:Hamiltonian-orig-withinRWA}
\end{equation}
Here, $\hat{\mathbf{v}}_k = (\hat{a}_k, \hat{b}_k)^T$ is the vector of annihilation operators at wavevector $k$ and $f(k)\equiv g_C(k) - g_L(k)$. In this case, the Hamiltonian is diagonalized to the form
\begin{equation}
    \hat{H} = \hbar\sum_{k}\left[ \omega_+(k)\:\hat{a}_{+,k}^\dagger \hat{a}_{+,k} + \omega_-(k)\:\hat{a}_{-,k}^\dagger \hat{a}_{-,k} \right],\label{eq:Hamiltonian-diagonalized-withinRWA}
\end{equation}
where two bands $\omega_{\pm}(k) = \omega_0(k) \pm |f(k)|$ symmetric with respect to $\omega_0(k)$ at each wavevector $k$ appear [here, note that $\hat{a}_{\pm,k}^\dagger \equiv (\hat{a}_{\pm,k})^\dagger$]. The supermodes $\hat{a}_{\pm,k}$ are written as
\begin{align*}
    \hat{a}_{\pm,k} = \frac{\pm e^{-i\phi(k)}\hat{a}_k +\hat{b}_k}{\sqrt{2}},
\end{align*}
where $\phi(k)\equiv\arg{f(k)}$ is the phase of coupling term. The Bloch states in the single-excitation bands are written as
\begin{equation*}
    |\psi_{k,\pm}\rangle = \hat{a}_{\pm,k}^\dagger |0\rangle = \frac{1}{\sqrt{2}}\left(\pm e^{i\phi(k)} |1_k,0_k\rangle + |0_k,1_k\rangle\right),
\end{equation*}

\noindent where $|n_k,n'_k\rangle$ denotes a state with $n$ ($n'$) photons in mode $\hat{a}_k$ ($\hat{b}_k$).

As discussed below in App.~\ref{sec:AppendixB}, the kernel of the Hamiltonian in Eq.~\eqref{eq:Hamiltonian-orig-withinRWA} has an inversion symmetry in the sublattice unit cell which is known to result in bands with quantized Zak phase~\cite{Zak:1989}. In our system the Zak phase of the two bands are evaluated as

\begin{align*}
\mathcal{Z} &= i\oint_{\text{B.Z.}}\ud k \: \frac{1}{\sqrt{2}}\begin{pmatrix} \pm e^{-i\phi(k)} & 1 \end{pmatrix} \frac{\partial}{\partial k} \left[\frac{1}{\sqrt{2}}\begin{pmatrix} \pm e^{i\phi(k)} \\ 1\end{pmatrix}\right] \\
&= -\frac{1}{2}\oint_{\text{B.Z.}} \ud k \: \frac{\partial \phi (k)}{\partial k}.
\end{align*}

\noindent The Zak phase of photonic bands is determined by the behavior of $f(k)$ in the complex plane. If the contour of $f(k)$ for $k$ values in the first Brillouin zone excludes (encloses) the origin, the Zak phase is given by $\mathcal{Z}=0$ ($\mathcal{Z}=\pi$) corresponding to the trivial (topological) phase.

\subsection{Band structure beyond the rotating-wave approximation}
Considering all the terms in the Hamiltonian in Eqs.~\eqref{eq:H0}-\eqref{eq:V}, the Hamiltonian can be written in a compact form $\hat{H} = \frac{\hbar}{2}\sum_{k}(\hat{\mathbb{v}}_k)^\dagger \mathbb{h}(k) \hat{\mathbb{v}}_k$ with a vector composed of mode operators $\hat{\mathbb{v}}_k = \left(\hat{a}_k,\hat{b}_k,   \hat{a}_{-k}^\dagger, \hat{b}_{-k}^\dagger \right)^T$ and
\begin{widetext}
\begin{align}
    {\mathbb{h}}(k) &=
    \begin{pmatrix}
    \omega_0 (k) & f(k) & 0 & g(k) \\
     f^*(k) & \omega_0(k) & g^*(k) & 0 \\
    0 & g(k) & \omega_0(k) & f(k) \\
    g^*(k) & 0 & f^*(k) & \omega_0(k)
    \end{pmatrix} 
    =
     \omega_0(k)
    \begin{pmatrix}
    1 & \frac{c_k - l_k}{2} & 0 & \frac{-c_k - l_k}{2} \\
    \frac{c_k^* - l_k^*}{2} & 1 & \frac{-c_k^* - l_k^*}{2} & 0 \\
    0 & \frac{-c_k - l_k}{2} & 1 & \frac{c_k - l_k}{2} \\
    \frac{-c_k^* - l_k^*}{2} & 0 & \frac{c_k^* - l_k^*}{2} & 1 
    \end{pmatrix},\label{eq:Hamiltonian-beyondRWA}
\end{align}
\end{widetext}
where $f(k) \equiv g_C(k)-g_L(k)$ as before and $g(k)\equiv -g_C(k)-g_L(k)$. Here, $l_k\equiv M_g(k)/L_0$ and $c_k\equiv C_g(k)/C_\Sigma$ are inductive and capacitive coupling normalized to frequency. The dispersion relation can be found by diagonalizing the kernel of the Hamiltonian in Eq.~\eqref{eq:Hamiltonian-beyondRWA} with the Bogoliubov transformation
\begin{equation}
\hat{\mathbb{w}}_k
 = \mathbf{S}_k \hat{\mathbb{v}}_k,\quad\quad \mathbf{S}_k= \begin{pmatrix}
 \mathbf{U}_{k} &  \mathbf{V}_{-k}^* \\
 \mathbf{V}_{k} & \mathbf{U}_{-k}^*
 \end{pmatrix}\label{eq:Sk-blocks}
\end{equation}
where $\hat{\mathbb{w}}_k\equiv (\hat{a}_{+,k},\hat{a}_{-,k}, \hat{a}_{+,-k}^\dagger,\hat{a}_{-,-k}^\dagger)^T$ is the vector composed of supermode operators and $\mathbf{U}_{k}$, $\mathbf{V}_{k}$ are $2\times 2$ matrices forming blocks in the transformation $\mathbf{S}_k$. We want to find $\mathbf{S}_k$ such that $(\hat{\mathbb{v}}_k)^\dagger \mathbb{h}(k) \hat{\mathbb{v}}_k = (\hat{\mathbb{w}}_k)^\dagger \Tilde{\mathbb{h}}(k) \hat{\mathbb{w}}_k$, where $\Tilde{\mathbb{h}}(k) $ is diagonal. To preserve the commutation relations, the matrix $\mathbf{S}_k$ has to be symplectic, satisfying $\mathbf{J} = \mathbf{S}_k\mathbf{J}(\mathbf{S}_k)^\dagger$, with $\mathbf{J}$ defined as \begin{equation*}
    \mathbf{J} = 
    \begin{bmatrix}
    1 & 0 & 0 & 0 \\
    0 & 1 & 0 & 0 \\
    0 & 0 & -1 & 0 \\
    0 & 0 & 0 & -1
    \end{bmatrix}.
\end{equation*}
Due to this symplecticity, it can be shown that the matrices $\mathbf{J}{\mathbb{h}}(k)$ and $\mathbf{J}\Tilde{\mathbb{h}}(k)$ are similar under transformation ${\mathbf{S}}_k$. Thus, finding the eigenvalues and eigenvectors of the coefficient matrix
\begin{equation}
    \mathbb{m}(k) \equiv \frac{\mathbf{J}\mathbb{h}(k)}{\omega_0(k)} = 
    \begin{pmatrix}
    1 & \frac{c_k - l_k}{2} & 0 & \frac{-c_k - l_k}{2} \\
    \frac{c_k^* - l_k^*}{2} & 1 & \frac{-c_k^* - l_k^*}{2} & 0 \\
    0 & \frac{c_k + l_k}{2} & -1 & \frac{-c_k + l_k}{2} \\
    \frac{c_k^* + l_k^*}{2} & 0 & \frac{-c_k^* + l_k^*}{2} & -1 
    \end{pmatrix}\label{eq:m-matrix}
\end{equation}
is sufficient to obtain the dispersion relation and supermodes of the system. The eigenvalues of matrix $\mathbb{m}(k)$ are evaluated as
\begin{widetext}
\begin{equation*}
 \pm \sqrt{1 - \frac{l_k c_k^* + 
    l_k^* c_k}{2} \pm \sqrt{\left(1 - \frac{l_k c_k^* + 
    l_k^* c_k}{2}\right)^2 - (1 - |l_k|^2)(1-|c_k|^2)}}
\end{equation*}
and hence the dispersion relation of the system taking into account all terms in Hamiltonian \eqref{eq:Hamiltonian-beyondRWA} is
\begin{align}
    \Tilde{\omega}_{\pm}(k) = \Tilde{\omega}_0(k)\sqrt{1 \pm \sqrt{1 - \frac{\left[L_0^2 - {M_g(-k)M_g(k)}\right]\left[C_\Sigma^2-{C_g(-k)C_g(k)}\right]}{\left\{L_0 C_\Sigma -\frac{1}{2}\left[{M_g(-k)C_g(k) + C_g(-k)M_g(k)}\right]\right\}^2}} }\label{eq:dispersion-relation-beyond-RWA}
\end{align}
where 
\begin{align*}
    \Tilde{\omega}_0(k) \equiv {\omega}_0(k)\sqrt{1-\frac{M_g(k)C_g(-k) + M_g(-k)C_g(k)}{2 L_0 C_\Sigma}}.
\end{align*}
The two passbands range over frequencies $[\omega_{+}^\text{min}, \omega_{+}^\text{max}]$ and $[\omega_{-}^\text{min}, \omega_{-}^\text{max}]$, where the band-edge frequencies are written as
\begin{subequations}
\begin{equation}
    \omega_{+}^\text{min}=\frac{1}{\sqrt{[L_0+p_2(M_v - M_w)][C_\Sigma - p_2(C_v-C_w)]}} \quad \text{and}\quad\omega_{+}^\text{max}=\frac{1}{\sqrt{[L_0+p_1(M_v+M_w)][C_\Sigma - p_1(C_v + C_w)]}},\label{eq:bandedgefreq-upper}
\end{equation}
\begin{equation}
    \omega_{-}^\text{min}=\frac{1}{\sqrt{[L_0 - p_1(M_v+M_w)][C_\Sigma + p_1(C_v + C_w)]}}\quad\text{and}\quad \omega_{-}^\text{max}=\frac{1}{\sqrt{[L_0-p_2(M_v - M_w)][C_\Sigma + p_2(C_v-C_w)]}}.\label{eq:bandedgefreq-lower}
\end{equation}
\end{subequations}
Here, $p_1\equiv \text{sgn}[L_0(C_v+C_w)-C_\Sigma(M_v+M_w)]$ and $p_2\equiv \text{sgn}[L_0(C_v-C_w)-C_\Sigma(M_v-M_w)]$ are sign factors.
In principle, the eigenvectors of the matrix $\mathbb{m}(k)$ in Eq.~\eqref{eq:m-matrix} can be analytically calculated to find the transformation $\mathbf{S}_k$ of the original modes to supermodes $\hat{a}_{\pm,k}$. For the sake of brevity, we perform the calculation in the limit of vanishing mutual inductance ($M_v=M_w=0$), where the matrix $\mathbb{m}(k)$ reduces to
\begin{equation}
    \mathbb{m}_C(k) \equiv  \begin{pmatrix}
    1 & c_k/2 & 0 & -c_k/2 \\
    c_k^*/2 & 1 & -c_k^*/{2} & 0 \\
    0 & c_k/2 & -1 & -c_k/2 \\
    c_k^*/{2} & 0 & -c_k^*/2 & -1 
    \end{pmatrix}.\label{eq:m-matrix-reduced}
\end{equation}
In this case, the block matrices $\mathbf{U}_{k}$, $\mathbf{V}_k$ in the transformation in Eq.~\eqref{eq:Sk-blocks} are written as
\begin{equation*}
\mathbf{U}_k = \frac{1}{2\sqrt{2}}
    \begin{pmatrix}
    e^{-i\phi(k)}x_{+,k} & x_{+,k} \\
    -e^{-i\phi(k)} x_{-,k} & x_{-,k}
    \end{pmatrix}
    ,\quad 
\mathbf{V}_k = \frac{1}{2\sqrt{2}}
    \begin{pmatrix}
    e^{-i\phi(k)}y_{+,k} & y_{+,k} \\
    -e^{-i\phi(k)} y_{-,k} & y_{-,k}
    \end{pmatrix},
\end{equation*}
where $x_{\pm,k}=\sqrt[4]{1\pm|c_k|}+ \frac{1}{\sqrt[4]{1\pm|c_k|}}$,  $y_{\pm,k}=\sqrt[4]{1\pm|c_k|}- \frac{1}{\sqrt[4]{1\pm|c_k|}}$, and $\phi(k)=\arg c_k$. Note that the constants are normalized by relation $x_{\pm,k}^2 - y_{\pm,k}^2 = 4$.

The knowledge of the transformation $\mathbf{S}_k$ allows us to evaluate the Zak phase of photonic bands. In the Bogoliubov transformation, the Zak phase can be evaluated as~\cite{Goren:2018}
\begin{align*}
\mathcal{Z} &= i\oint_{\text{B.Z.}}\ud k \: \frac{1}{2\sqrt{2}}\begin{pmatrix} \pm e^{-i\phi(k)} x_{\pm,k} & x_{\pm,k}  & \pm e^{-i\phi(k)} y_{\pm,k} & y_{\pm,k}  \end{pmatrix} \cdot \mathbf{J} \cdot \frac{\partial}{\partial k} \left[\frac{1}{2\sqrt{2}}\begin{pmatrix} \pm e^{i\phi(k)}  x_{\pm,k}  \\  x_{\pm,k} \\ \pm e^{i\phi(k)}  y_{\pm,k}   \\ y_{\pm,k} \end{pmatrix}\right] \nonumber \\
&= i\oint_{\text{B.Z.}}\ud k\: \frac{1}{8} \left[i\frac{\partial\phi(k)}{\partial k} (x_{\pm,k}^2 - y_{\pm,k}^2) + \frac{\partial}{\partial k}(x_{\pm,k}^2 - y_{\pm,k}^2)\right]
= -\frac{1}{2}\oint_{\text{B.Z.}} \ud k \: \frac{\partial \phi (k)}{\partial k},
\end{align*}
identical to the expression within the RWA. Again, the Zak phase of photonic bands is determined by the winding of $f(k)$ around the origin in complex plane, leading to $\mathcal{Z} = 0$ in the trivial phase and $\mathcal{Z} = \pi$ in the topological phase.
\end{widetext}

\subsection{Extraction of circuit parameters and the breakdown of the circuit model}
As discussed in Fig.~\ref{fig:Figure1}d of the main text, the parameters in the circuit model of the topological waveguide is found by fitting the waveguide transmission spectrum of the test structures. We find that two lowest-frequency modes inside the lower passband fail to be captured according to our model with capacitively and inductively coupled LC resonators. We believe that this is due to the broad range of frequencies (about 1.5\:GHz) covered in the spectrum compared to the bare resonator frequency $\sim6.6\:\text{GHz}$ and the distributed nature of the coupling, which can cause our simple model based on frequency-independent lumped elements (inductor, capacitor, and mutual inductance) to break down. Such deviation is also observed in the fitting of waveguide transmission data of Device I (Fig.~\ref{fig:FigureS7}).

\section{Mapping of the system to the SSH model and discussion on robustness of edge modes}
\label{sec:AppendixB}
\subsection{Mapping of the topological waveguide to the SSH model}
We discuss how the physical model of topological waveguide in App.~\ref{sec:AppendixA} could be mapped to the photonic SSH model, whose Hamiltonian is given as Eq.~\eqref{eq:Hamiltonian-SSH} in the main text. Throughout this section, we consider the realistic circuit parameters extracted from fitting of test structures given in Fig.~\ref{fig:Figure1} of the main text: resonator inductance and resonator capacitance, $L_0 = 1.9\:\text{nH}$ and $C_0=253\:\text{fF}$, and coupling capacitance and parasitic mutual inductance, $(C_v, C_w)=(33, 17)\:\text{fF}$ and $(M_v, M_w)=(-38, -32)\:\text{pH}$ in the trivial phase (the values are interchanged in the topological phase). 

\begin{figure}[b]
\centering
\includegraphics[scale=0.45]{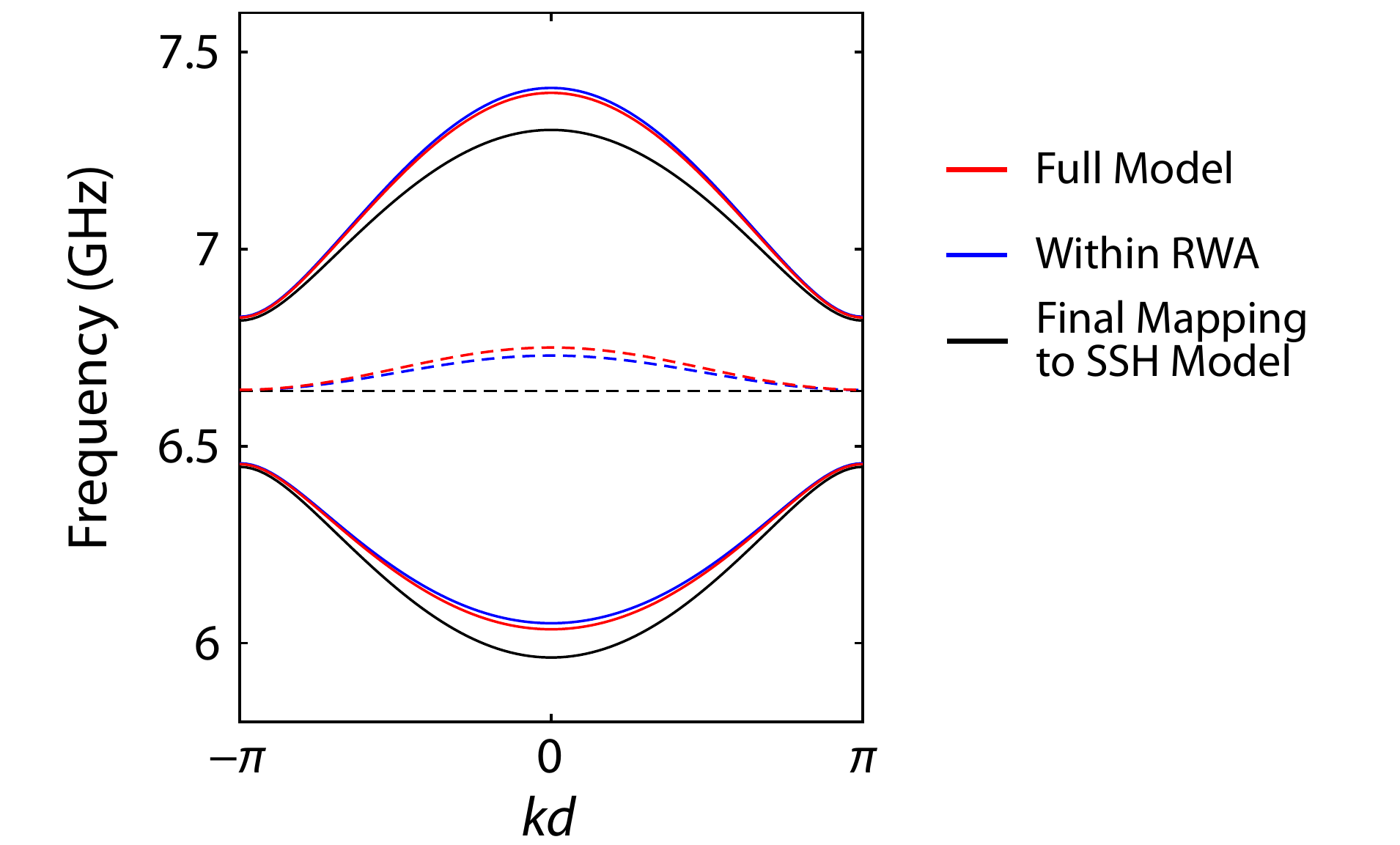}
  \caption{Band structure of the realized topological waveguide under various assumptions discussed in App.~\ref{sec:AppendixB}. The solid lines show the dispersion relation in the upper (lower) passband, $\omega_\pm(k)$: full model without any assumptions (red), model within RWA (blue), and the final mapping to SSH model (black) in the weak coupling limit. The dashed lines indicate the uncoupled resonator frequency $\omega_0(k)$ under corresponding assumptions.}
  \label{fig:FigureS2}
\end{figure}

To most directly and simply link the Hamiltonian described in Eqs.~\eqref{eq:H0}-\eqref{eq:V} to the SSH model, here we impose a few approximations. First, the counter-rotating terms in the Hamiltonian are discarded such that only photon-number-conserving terms are left. To achieve this, the RWA is applied to reduce the kernel of the Hamiltonian into one involving a $2\times 2$ matrix as in Eq.~\eqref{eq:Hamiltonian-orig-withinRWA}. Such an assumption is known to be valid when the coupling terms in the Hamiltonian are much smaller than the frequency scale of the uncoupled Hamiltonian $\hat{H}_0$~\cite{Kockum:2019}. According to the coupling terms derived in Eq.~\eqref{eq:coupling-terms}, this is a valid approximation given that
\begin{align*}
\left|\frac{g_C(k)}{\omega_0(k)}\right|&\le\frac{|C_v+C_w|}{2C_\Sigma}\approx 0.083,\\ \left|\frac{g_L(k)}{\omega_0(k)}\right|&\le\frac{|M_v|+|M_w|}{2L_0}\approx 0.018.
\end{align*}
and the RWA affects the dispersion relation by less than 0.3\:\% in frequency. 

Also different than in the original SSH Hamiltonian, are the $k$-dependent diagonal elements $\omega_0(k)$ of the single-particle kernel of the Hamiltonian for the circuit model. This $k$-dependence can be understood as arising from the coupling between resonators beyond nearest-neighbor pairs, which is inherent in the canonical quantization of capacitively coupled LC resonator array (due to circuit topology) as discussed in Ref.~\cite{Ferreira:2020}. The variation in $\omega_0(k)$ can be effectively suppressed in the limit of $C_v, C_w \ll C_\Sigma$ and $M_v, M_w\ll L_0$ as derived in Eq.~\eqref{eq:effective-cap-ind}. We note that while our coupling capacitances are small compared to  $C_\Sigma$ ($C_v/C_\Sigma \approx 0.109$, $C_w/C_\Sigma \approx 0.056$ in the trivial phase), we find that they are sufficient to cause the $\omega_0(k)$ to vary by $\sim$1.2\:\% in the first Brillouin zone. Considering this limit of small coupling capacitance and mutual inductance, the effective capacitance and inductance of \eqref{eq:effective-cap-ind} become quantities independent of $k$, $C_0^\text{eff}(k)\approx C_\Sigma$, $L_0^\text{eff}(k)\approx L_0$, and the kernel of the Hamiltonian under RWA reduces to

\begin{equation*}
    \mathbf{h}(k)=
    \begin{pmatrix}
    \omega_0 & f(k) \\
    f^*(k) & \omega_0
    \end{pmatrix}.
\end{equation*}
Here, 
\begin{widetext}
\begin{equation*}
    \omega_0 = \frac{1}{\sqrt{L_0C_\Sigma}},\quad f(k)=\frac{\omega_0}{2}\left[ \left( \frac{C_v}{C_\Sigma} - \frac{M_v}{L_0}\right) + \left(\frac{C_w}{C_\Sigma} - \frac{M_w}{L_0}\right)e^{-ikd}\right].
\end{equation*}

\noindent This is equivalent to the photonic SSH Hamiltonian in Eq.~\eqref{eq:Hamiltonian-SSH} of the main text under redefinition of gauge which transforms operators as  $(\hat{a}_k,\hat{b}_k)\rightarrow (\hat{a}_k,-\hat{b}_k)$. Here, we can identify the parameters $J$ and $\delta$ as
\begin{equation}
    J = \frac{\omega_0}{4}\left(\frac{C_v + C_w}{C_\Sigma} - \frac{M_v + M_w}{L_0}\right),\quad \delta = \frac{L_0(C_v - C_w) - {C_\Sigma}({M_v - M_w})}{L_0({C_v + C_w}) - {C_\Sigma}({M_v + M_w})},\label{eq:J-delta}
\end{equation}
\end{widetext}
where $J(1\pm \delta)$ is defined as intra-cell and inter-cell coupling, respectively. The dispersion relations under different stages of approximations mentioned above are plotted in Fig.~\ref{fig:FigureS2}, where we find a clear deviation of our system from the original SSH model due to the $k$-dependent reference frequency.

\subsection{Robustness of edge modes under perturbation in circuit parameters}
While we have linked our system to the SSH Hamiltonian in Eq.~\eqref{eq:Hamiltonian-SSH} of the main text, we find that our system fails to strictly satisfy chiral symmetry $\mathcal{C}{\mathbf{h}}(k)\mathcal{C}^{-1}=-\mathbf{h}(k)$ ($\mathcal{C} = \sigmaz$ is the chiral symmetry operator in the sublattice space).  This is due to the $k$-dependent diagonal $\omega_0(k)$ terms in $\mathbf{h}(k)$, resulting from the non-local nature of the quantized charge and nodal flux in the circuit model which results in next-nearest-neighbor coupling terms between sublattices of the same type.  Despite this, an inversion symmetry, $\mathcal{I}{\mathbf{h}}(k)\mathcal{I}^{-1}={\mathbf{h}}(-k)$ ($\mathcal{I}=\sigmax$ in the sublattice space), still holds for the circuit model.  This ensures the quantization of the Zak phase ($\mathcal{Z}$) and the existence of an invariant band winding number ($\nu=\mathcal{Z}/\pi$) for perturbations that maintain the inversion symmetry.  However, as shown in Refs.~\cite{Perez-Gonzalez:2018,Longhi:2018}, the inversion symmetry does not protect the edge states for highly delocalized coupling along the dimer resonator chain, and the correspondence between winding number and the number of localized edge states at the boundary of a finite section of waveguide is not guaranteed.

 \begin{figure*}[t!]
\centering
\includegraphics[width=0.65\textwidth]{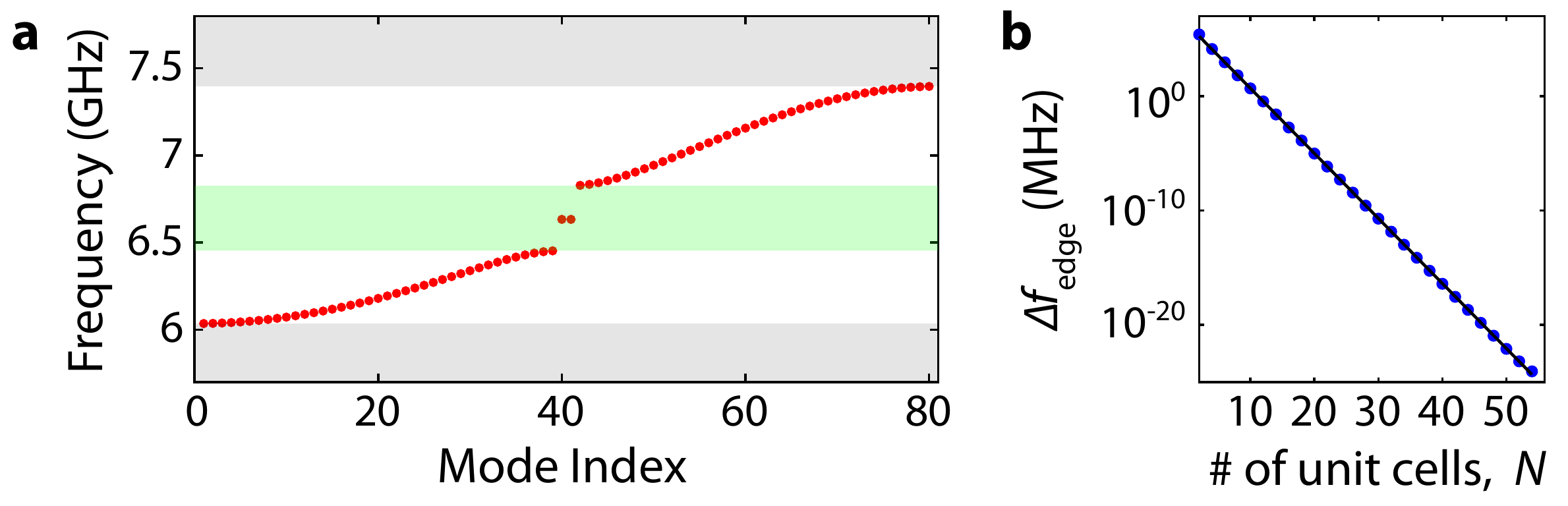}
  \caption{\textbf{a}, Resonant frequencies of a finite system with $N=40$ unit cells, calculated from eigenmodes of Eq.~\eqref{eq:classical-eigeneq}. The bandgap regions calculated from dispersion relation are shaded in gray (green) for upper and lower bandgaps (middle bandgap). The two data points inside the middle bandgap (mode indices 40 and 41) correspond to edge modes. \textbf{b}, Frequency splitting $\Delta f_\text{edge}$ of edge modes with no disorder in the system are plotted against the of number of unit cells $N$. The black solid curve indicates exponential fit to the edge mode splitting, with decay constant of $\xi=1.76$.}
  \label{fig:FigureS3}
\end{figure*}
 

For weak breaking of the chiral symmetry (i.e., beyond nearest-neighbor coupling much smaller than nearest neighbor coupling) the correspondence between winding number and the number of pairs of gapped edge states is preserved, with winding number $\nu=0$ in the trivial phase ($\delta > 0$) and $\nu=1$ in the topological ($\delta < 0$) phase.  Beyond just the existence of the edge states and their locality at the boundaries, chiral symmetry is special in that it pins the edge mode frequencies at the center of the middle bandgap ($\omega_0$).  Chiral symmetry is maintained in the presence of disorder in the coupling between the different sublattice types along the chain, providing stability to the frequency of the edge modes.  In order to study the robustness of the edge mode frequencies in our circuit model, we perform a simulation over different types of disorder realizations in the circuit illustrated in Fig.~\ref{fig:FigureS1}. As the original SSH Hamiltonian with chiral symmetry gives rise to topological edge states which are robust against the disorder in coupling, not in on-site energies~\cite{Asboth:2016}, it is natural to consider disorder in circuit elements that induce coupling between resonators: $C_v$, $C_w$, $M_v$, $M_w$. 

The classical equations of motion of a circuit consisting of $N$ unit cells is written as
\begin{widetext}
\begin{align*}
    V_n^\text{A} &= L_0 \frac{\ud{i}_n^\text{A}}{\ud t} + M_v^{(n)} \frac{\ud{i}_n^\text{B}}{\ud t} + M_w^{(n)} \frac{\ud{i}_{n-1}^\text{B}}{\ud t}, \quad
    i_n^\text{A} = -C_{\Sigma,\text{A}}^{(n)} \frac{\ud{V}_n^\text{A}}{\ud t} + C_v^{(n)} \frac{\ud{V}_n^\text{B}}{\ud t} + C_w^{(n-1)} \frac{\ud{V}_{n-1}^\text{B}}{\ud t}\\
    V_n^\text{B} &= L_0 \frac{\ud{i}_n^\text{B}}{\ud t} + M_w^{(n)}, \frac{\ud{i}_{n+1}^\text{A}}{\ud t} + M_v^{(n)} \frac{\ud{i}_{n}^\text{A}}{\ud t},\quad
    i_n^\text{B} = -C_{\Sigma,\text{B}}^{(n)} \frac{\ud{V}_n^\text{B}}{\ud t} + C_v^{(n)} \frac{\ud{V}_n^\text{A}}{\ud t} + C_w^{(n)} \frac{\ud{V}_{n+1}^\text{A}}{\ud t},
\end{align*}
where the superscripts indicate index of cell of each circuit element and
\begin{equation*}
    C_{\Sigma,\text{A}}^{(n)} = C_0 + C_v^{(n)} + C_w^{(n-1)},\quad    C_{\Sigma,\text{B}}^{(n)} = C_0 + C_v^{(n)} + C_w^{(n)}.
\end{equation*}
The $4N$ coupled differential equations are rewritten in a compact form as
\begin{equation}
    \frac{\ud}{\ud t}
    \begin{pmatrix}
    \mathbf{u}_1 \\
    \mathbf{u}_2 \\
    \vdots \\
    \mathbf{u}_N
    \end{pmatrix} = \mathbf{C}^{-1} 
    \begin{pmatrix}
    \mathbf{u}_1 \\
    \mathbf{u}_2 \\
    \vdots \\
    \mathbf{u}_N
    \end{pmatrix} ,
    \quad
    \mathbf{u}_{n} \equiv
    \begin{pmatrix}
    V_n^\text{A} \\ i_n^\text{A} \\ V_n^\text{B} \\ i_n^\text{B}
    \end{pmatrix},\label{eq:classical-eigeneq}
\end{equation}
where the coefficient matrix $\mathbf{C}$ is given by
\setcounter{MaxMatrixCols}{19}
\begin{align*}
\mathbf{C} \equiv
    \begin{pmatrix}
     0 & L_0 & 0 & M_v^{(1)} &  &  &  &  &  &  & & & &  \\
    -C_{\Sigma,\text{A}}^{(1)} & 0 & C_v^{(1)} & & & &  &  &  & & & & & \\
     & M_v^{(1)} & 0 & L_0 & 0 & M_w^{(1)} & & &    &  &  & & & \\
     C_v^{(1)} & 0 & -C_{\Sigma,\text{B}}^{(1)} & 0 & C_w^{(1)} &  & & & &  &  & & &  \\
     & & & M_w^{(1)} & 0 & L_0 & 0 & M_v^{(2)} & &  & & & & \\
     & & C_w^{(1)} & 0 & -C_{\Sigma,\text{A}}^{(2)} & 0 & C_v^{(2)}   & &  & & & & &  \\
     & & & & & M_v^{(2)} & 0 & L_0 & 0 & M_w^{(2)}  &  &  & & & &  \\
     & & & & C_v^{(2)} & 0 & -C_{\Sigma,\text{B}}^{(2)} & 0 & C_w^{(2)} &  & \ddots & & &  \\
     & & & & & & & M_w^{(2)} & & \ddots &  &\ddots & &  \\
     & & & & & & C_w^{(2)} &  & \ddots & & \ddots &  & \ddots &  \\
     & & & & & & & \ddots &  & \ddots &  & \ddots &    &  M_v^{(N)}   \\
     & & & & & & &  & \ddots &  & \ddots &  &  C_v^{(N)}  &   \\
     & & & & & & & & & \ddots &  & M_v^{(N)} & 0 & L_0    \\
         & & & & & & & &   &  & C_v^{(N)} & 0 & -C_{\Sigma,\text{B}}^{(N)} &  0 \\
    \end{pmatrix}.
\end{align*}
\end{widetext}
Here, the matrix elements not specified are all zero. The resonant frequencies of the system can be determined by finding the positive eigenvalues of $i \mathbf{C}^{-1}$. Considering the model without any disorder, we find the eigenfrequencies of the finite system to be distributed according to the passband and bandgap frequencies from dispersion relation in Eq.~\eqref{eq:dispersion-relation-beyond-RWA}, as illustrated in Fig.~\ref{fig:FigureS3}. Also, we observe the  presence of a pair of coupled edge mode resonances inside the middle bandgap in the topological phase, whose splitting due to finite system size scales as $\Delta f_\text{edge}\sim e^{-N/\xi}$ with $\xi = 1.76$.

\begin{figure*}[ht!]
\centering
\includegraphics[width=0.8\textwidth]{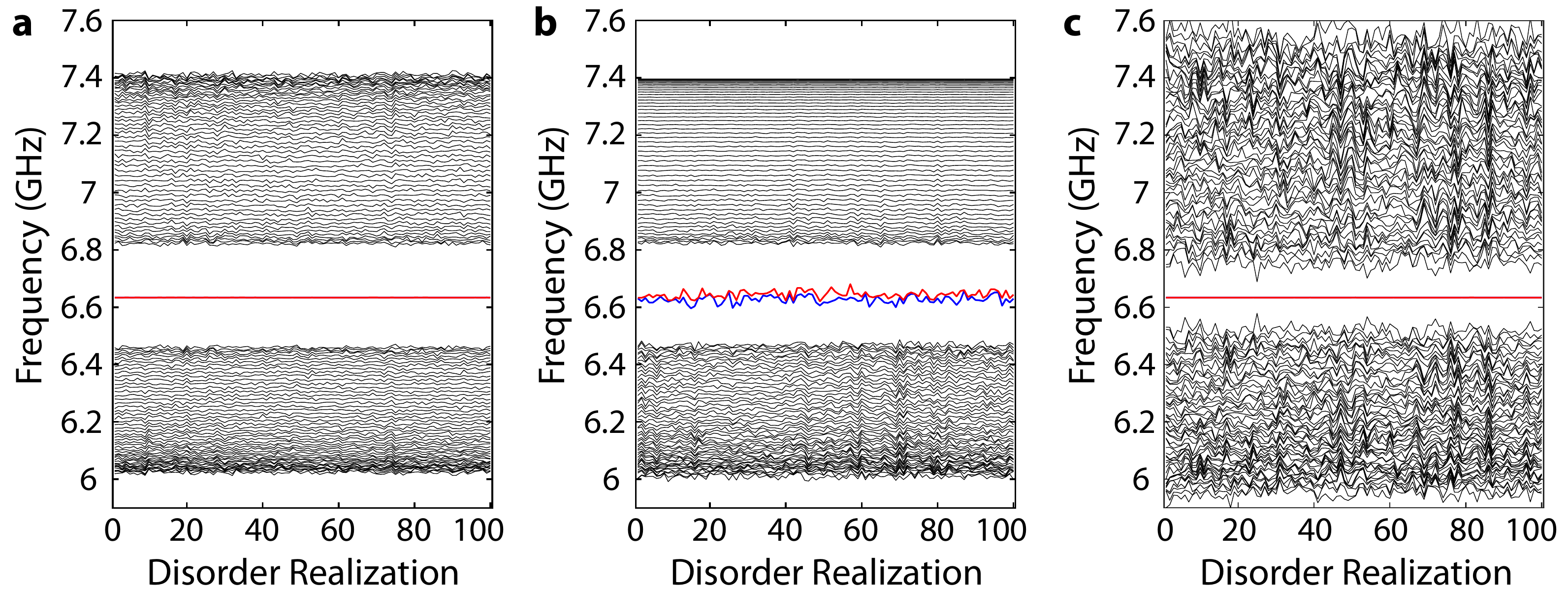}
  \caption{\textbf{Eigenfrequencies of the system under 100 disorder realizations in coupling elements.} Each disorder realization is achieved by uniformly sampling the parameters within fraction $\pm r$ of the original value. \textbf{a}, Disorder in mutual inductance $M_v$ and $M_w$ between neighboring resonators with the strength $r=0.5$. \textbf{b}, Disorder in coupling capacitance $C_v$ and $C_w$ between neighboring resonators with the strength $r=0.1$. \textbf{c}, The same disorder as panel \textbf{b} with $r=0.5$, while keeping the bare self-capacitance $C_\Sigma$ of each resonator fixed (correlated disorder between coupling capacitances and resonator $C_0$).}
  \label{fig:FigureS4}
\end{figure*}

To discuss the topological protection of the edge modes, we keep track of the set of eigenfrequencies for different disorder realizations of the coupling capacitance and mutual inductance for a system with $N=50$ unit cells. First, we consider the case when the mutual inductance $M_v$ and $M_w$ between resonators are subject to disorder. The values of $M_v^{(n)}$, $M_w^{(n)}$ are assumed to be sampled uniformly on an interval covering a fraction $\pm r$ of the original values, i.e.,
$$
M_v^{(n)} = M_v \left[ 1 + r \tilde{\delta}^{(n)}_{M_v}\right],\quad M_w^{(n)} = M_w \left[1 + r \tilde{\delta}^{(n)}_{M_w}\right],
$$
where $\tilde{\delta}^{(n)}_{M_v},\tilde{\delta}^{(n)}_{M_w}$ are independent random numbers uniformly sampled from an interval $[-1,1]$. Figure~\ref{fig:FigureS4}a illustrates an example with a strong disorder with $r=0.5$ under 100 independent realizations, where we find the frequencies of the edge modes to be stable, while frequencies of modes in the passbands fluctuate to a much larger extent. This suggests that the frequencies of edge modes have some sort of added robustness against disorder in the mutual inductance between neighboring resonators despite the fact that our circuit model does not satisfy chiral symmetry. The reduction in sensitivity results from the fact that the effective self-inductance $L_0^\text{eff}(k)$ of the resonators, which influences the on-site resonator frequency, depends on the mutual inductances only to second-order in small parameter $(M_{v,w}/L_0)$.  It is this second-order fluctuation in the resonator frequencies, causing shifts in the diagonal elements of the Hamiltonian, which results in fluctuations in the edge mode frequencies. The direct fluctuation in the mutual inductance couplings themselves, corresponding to off-diagonal Hamiltonian elements, do not cause the edge modes to fluctuate due to chiral symmetry protection (the off-diagonal part of the kernel of the Hamiltonian is chiral symmetric).

Disorder in coupling capacitance $C_v$ and $C_w$ are also investigated using a similar model, where the values of $C_v^{(n)}$, $C_w^{(n)}$ are allowed to vary by a fraction $\pm r$ of the original values (uniformly sampled), while the remaining circuit parameters are kept constant. From Fig.~\ref{fig:FigureS4}b we observe severe fluctuations in the frequencies of the edge modes even under a mild disorder level of $r=0.1$. This is due to the fact that the coupling capacitance $C_v$ and $C_w$ contribute to the effective self-capacitance of each resonator $C_0^\text{eff}(k)$ to first-order in small parameter $(C_{v,w}/C_0)$, thus directly breaking chiral symmetry and causing the edge modes to fluctuate.  An interesting observation in Fig.~\ref{fig:FigureS4}b is the stability of frequencies of modes in the upper passband with respect to disorder in $C_v$ and $C_w$. This can be explained by noting the expressions for band-edge frequencies in Eqs.~\eqref{eq:bandedgefreq-upper}-\eqref{eq:bandedgefreq-lower}, where the dependence on coupling capacitance gets weaker close to the upper band-edge frequency $\omega_+^\text{max}=1/\sqrt{(L_0+M_v+M_w)C_0}$ of the upper passband. 

Finally, we consider a special type of disorder where we keep the bare self-capacitance $C_\Sigma$ of each resonator fixed. Although unrealistic, we allow $C_v$ and $C_w$ to fluctuate and compensate for the disorder in $C_\Sigma$ by subtracting the deviation in $C_v$ and $C_w$ from $C_0$. This suppresses the lowest-order resonator frequency fluctuations, and hence helps stabilize the edge mode frequencies even under strong disorder $r=0.5$, as illustrated in Fig.~\ref{fig:FigureS4}c. While being an unrealistic model for disorder in our physical system, this observation sheds light on the fact that the circuit must be carefully designed to take advantage of the topological protection. It should also be noted that in all of the above examples, the standard deviation in the edge mode frequencies scale linearly to lowest order with the standard deviation of the disorder in the inter- and intra-cell coupling circuit elements (only the pre-coefficient changes).  Exponential suppression of edge mode fluctuations due to disorder in the coupling elements as afforded by the SSH model with chiral symmetry would require a redesign of the circuit to eliminate the next-nearest-neighbor coupling present in the current circuit layout.   

\section{Device I characterization and Experimental setup}
\label{sec:AppendixC}
\begin{figure*}[t!]
\centering
\includegraphics[width = \textwidth]{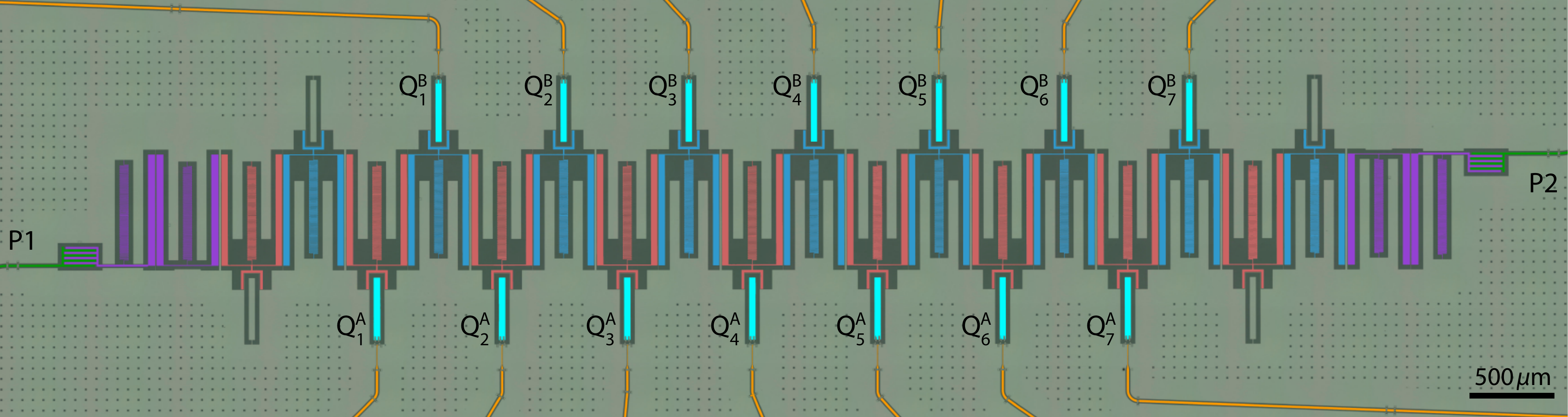}
  \caption{\textbf{Optical micrograph of Device I (false-colored).} The device consists of a topological waveguide with 9 unit cells (resonators corresponding to A/B sublattice colored red/blue) in the trivial phase, where the intra-cell coupling is larger than the inter-cell coupling. Qubits (cyan, labeled Q$_j^\alpha$ where $i$=1-7 and $\alpha$=A,B) are coupled to every site of the seven inner unit cells of the topological waveguide, each connected to on-chip flux-bias lines (orange) for individual frequency control. At the boundary of the topological waveguide are tapering sections (purple), which provide impedance matching to the external waveguides (green) at upper bandgap frequencies. P1 (P2) denotes port 1 (port 2) of the device.
  }
  \label{fig:FigureS5}
\end{figure*}

In this section, we provide a detailed description of elements on Device I, where the directional qubit-photon bound state and passband topology experiments are performed. The optical micrograph of Device I is shown in Fig.~\ref{fig:FigureS5}.

\begin{table*}[t!]
\begin{threeparttable}[t]
  \centering
\begin{tabular}{M{1.8cm}P{1.0cm}P{1.0cm}P{1.0cm}P{1.0cm}P{1.0cm}P{1.0cm}P{1.0cm}P{1.0cm}P{1.0cm}P{1.0cm}P{1.0cm}P{1.0cm}P{1.0cm}P{1.0cm}} 
               & Q$_1^\text{A}$    & Q$_1^\text{B}$    & Q$_2^\text{A}$    & Q$_2^\text{B}$  & Q$_3^\text{A}$    & Q$_3^\text{B}$    & Q$_4^\text{A}$ & Q$_4^\text{B}$ & Q$_5^\text{A}$ & Q$_5^\text{B}$ & Q$_6^\text{A}$ & Q$_6^\text{B}$ & Q$_7^\text{A}$ & Q$_7^\text{B}$   \\ \hline \hline
$\Gamma^\prime/2\pi$  (kHz)  &    325.7   &    150.4   &  247.4   &    104.7\tnote{a}   &     268.2  &    183.2   &  220.6 & 224.4 & 193.3 & 263.2 & 206 & 332.69 & 88.1 & 346.8
 \end{tabular}
     \begin{tablenotes}
     \item[a]Measured in a separate cooldown
   \end{tablenotes}
   \end{threeparttable}
   \caption{\textbf{Qubit coherence in the middle bandgap.} The parasitic decoherence rate $\Gamma^\prime$ of qubits on Device I at 6.621 GHz inside the MBG. The data for Q$_2^\text{B}$ was taken in a separate cooldown due to coupling to a two-level system defect.}
   \label{tb:qubits}
\end{table*}
\subsection{Qubits}
All 14 qubits on Device I are designed to be nominally identical with asymmetric Josephson junctions (JJs) on superconducting quantum interference device (SQUID) loop to reduce the sensitivity to flux noise away from maximum and minimum frequencies, referred to as ``sweet spots''. The sweet spots of all qubits lie deep inside the upper and lower bandgaps, where the coupling of qubits to external ports are small due to strong localization. This makes it challenging to access the qubits with direct spectroscopic methods near the sweet spots. Alternatively, a strong drive tone near resonance with a given qubit frequency was sent into the waveguide to excite the qubit, and a passband mode dispersively coupled to the qubit is simultaneously monitored with a second probe tone. With this method, the lower (upper) sweet spot of Q$_1^\text{A}$ is found to be at 5.22\:GHz (8.38\:GHz), and the anharmonicity near the upper sweet spot is measured to be 297\:MHz (effective qubit capacitance of $C_\text{q} = 65\:\text{fF}$). The Josephson energies of two JJs of Q$_1^\text{A}$ are extracted to be $(E_{J1}, E_{J2})/h = (21.85, 9.26)\:\text{GHz}$ giving the junction asymmetry of $d = \frac{E_{J1} - E_{J2}}{E_{J1} + E_{J2}} = 0.405$.

The coherence of qubits is characterized using spectroscopy inside the middle bandgap (MBG). Here, the parasitic decoherence rate is defined as $\Gamma' \equiv 2\Gamma_2 - \kappa_{e,1} - \kappa_{e,2}$, where $2\Gamma_2$ is the total linewidth of qubit, and $\kappa_{e,1}$ ($\kappa_{e,2}$) is the external coupling rate to port 1 (2) (see Supplementary Note 1 of Ref.~\cite{Mirhosseini:2019} for a detailed discussion). Here, $\Gamma'$ contains contributions from both qubit decay to spurious channels other than the desired external waveguide as well as pure dephasing. Table~\ref{tb:qubits} shows the parasitic decoherence rate of all 14 qubits at 6.621\:GHz extracted from spectroscopic measurement at a power at least 5\:dB below the single-photon level (defined as $\hbar \omega \kappa_{e,p}$ with $p = 1,2$) from both ports.

Utilizing the dispersive coupling between the qubit and a resonator mode in the passband, we have also performed  time-domain characterization of qubits. The measurement on Q$_4^\text{B}$ at 6.605 GHz in the MBG gives $T_1=1.23\:\mu\text{s}$ and $T_2^*=783\:\text{ns}$ corresponding to $\Gamma'/2\pi=281.3\:\text{kHz}$, consistent with the result from spectroscopy in Table~\ref{tb:qubits}. At the upper sweet spot, Q$_4^\text{B}$ was hard to access due to the small coupling to external ports arising from short localization length and a large physical distance from the external ports. Instead, Q$_1^\text{B}$ is characterized to be  $T_1= 9.197\:\mu\text{s}$ and $T_2^*=11.57\:\mu\text{s}$ at its upper sweet spot (8.569\:GHz).

\begin{figure*}[ht!]
\centering
\includegraphics[width = \textwidth]{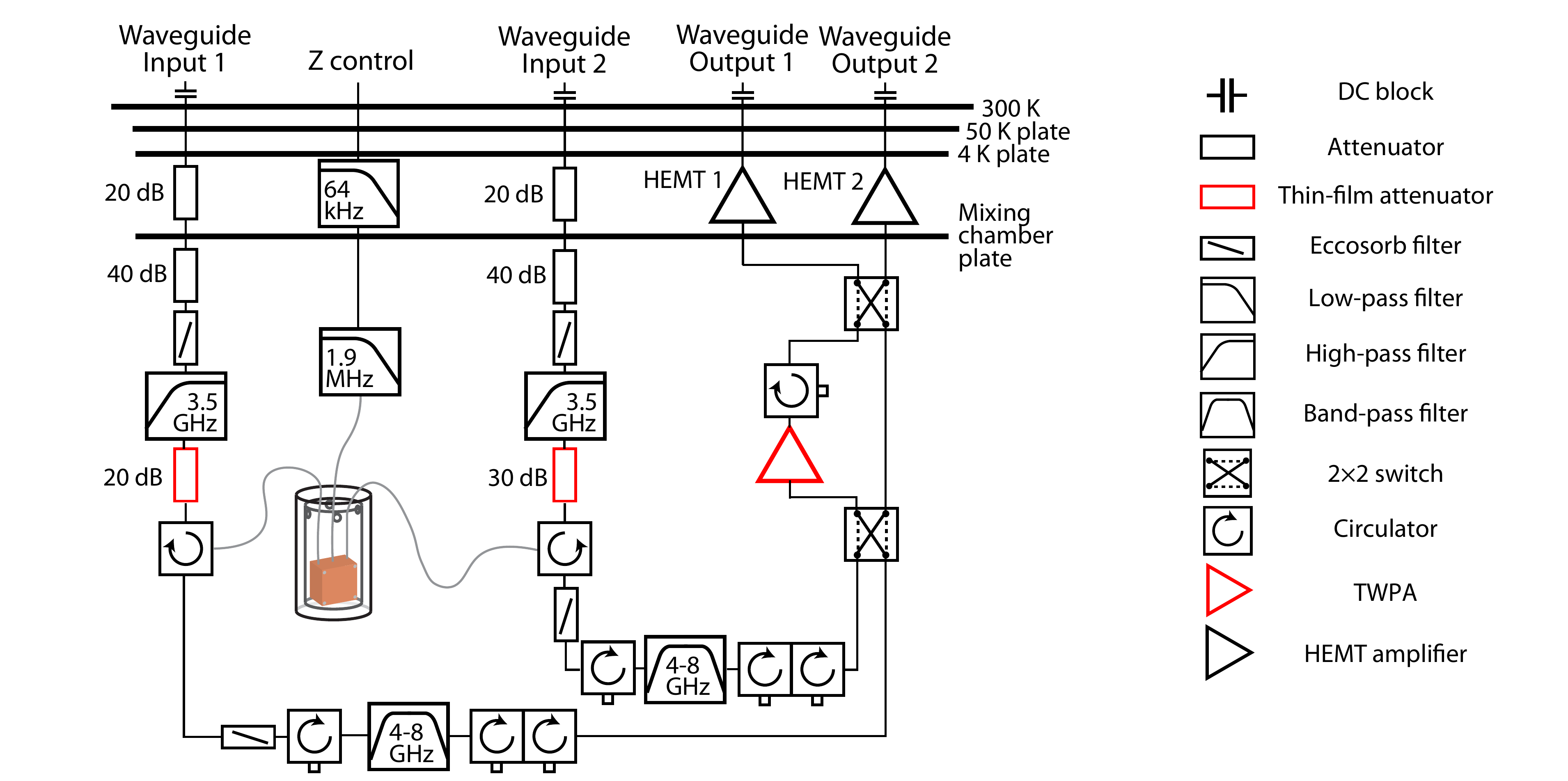}
  \caption{\textbf{Schematic of the measurement setup inside the dilution refrigerator for Device I.} The meaning of each symbol in the schematic on the left is enumerated on the right. The level of attenuation of each attenuator is indicated with number next to the symbol. The cutoff frequencies of each filter is specified with numbers inside the symbol. Small squares attached to circulator symbols indicate port termination with $Z_0=50\:\Omega$, allowing us to use the 3-port circulator as a 2-port isolator. The input pump line for TWPA is not shown in the diagram for simplicity.}
  \label{fig:FigureS6}
\end{figure*}

\subsection{Metamaterial waveguide and coupling to qubits}

As shown in Fig.~\ref{fig:FigureS5}, the metamaterial waveguide consists of a SSH array in the trivial configuration and tapering sections at the boundary (the design of tapering sections is discussed in App.~\ref{sec:AppendixD}). The array contains 18 identical LC resonators, whose design is slightly different from the one in test structures shown in Fig.~\ref{fig:Figure1}b of the main text. Namely, the ``claw'' used to couple qubits to resonators on each site is extended to generate a larger coupling capacitance of $C_g=5.6\:\text{fF}$ and  the resonator capacitance to ground was reduced accordingly to maintain the designed reference frequency. On resonator sites where no qubit is present, an island with shape identical to that of a qubit was patterned and shorted to ground plane in order to mimic the self-capacitance contribution from a qubit to the resonator. The fitting of the whole structure to the waveguide transmission spectrum results in a set of circuit parameters similar yet slightly different from ones of the test structures quoted in Fig.~\ref{fig:Figure1} of the main text: $(C_v, C_w) = (35, 19.2)\:\text{fF}$, $(M_v, M_w) = (-38, -32)\:\text{pH}$, $C_0 = 250\:\text{fF}$, $L_0 = 1.9\:\text{nH}$. Here, the definition of $C_0$ includes contributions from coupling capacitance between qubit and resonator, but excludes the contribution to the resonator self-capacitance from the coupling capacitances $C_v$, $C_w$ between resonators in the array. With these parameters we calculate the corresponding parameters in the SSH model to be $J/2\pi=356\:\text{MHz}$ and $\delta=0.256$ following Eq.~\eqref{eq:J-delta}, resulting in the localization length $\xi = [\ln(\frac{1+\delta}{1-\delta})]^{-1}=1.91$ at the reference frequency. From the measured avoided crossing $g_{45}^\text{AB}/2\pi = 32.9\:\text{MHz}$ between qubit-photon bound states facing toward each other on nearest-neighboring sites together with $J$ and $\delta$, we infer the qubit coupling to each resonator site to be $g = \sqrt{g_{45}^\text{AB} J(1+\delta)} = 2\pi\times 121.3\:\text{MHz}$~\cite{Bello:2019}, close to the value  $$\frac{C_g}{2\sqrt{C_\text{q} C_\Sigma}}\omega_0 = 2\pi\times 132\:\text{MHz}$$
expected from designed coupling capacitance~\cite{Sank:2014}. Note that we find an inconsistent set of values $J/2\pi = 368\:\text{MHz}$ and $\delta = 0.282$ (with $\xi=1.73$ and $g/2\pi=124.6\:\text{MHz}$ accordingly) from calculation based on the difference in observed band-edge frequencies, where the frequency difference between the highest frequency in the UPB and the lowest frequency in the LPB equals $4J$ and the size of the MBG equals $4J|\delta|$. The inconsistency indicates the deviation of our system from the proposed circuit model (see App.~\ref{sec:AppendixA} for discussion), which accounts for the difference between theoretical curves and the experimental data in Fig.~\ref{fig:Figure1}d and left sub-panel of Fig.~\ref{fig:Figure2}c. The values of $J, \delta$ and $g$ from the band-edge frequencies are used to generate the theoretical curves in Fig.~\ref{fig:Figure3} in the main text as well as in Fig.~\ref{fig:FigureS11}. The intrinsic quality factor of one of the normal modes (resonant frequency 6.158\:GHz) of the metamaterial waveguide was measured to be $Q_i= 9.8 \times 10^{\text{4}}$ at power below the single-photon level, similar to typical values reported in Refs.~\cite{Mirhosseini:2018, Ferreira:2020}.

\begin{figure*}[ht!]
\centering
\includegraphics[width = 0.9\textwidth]{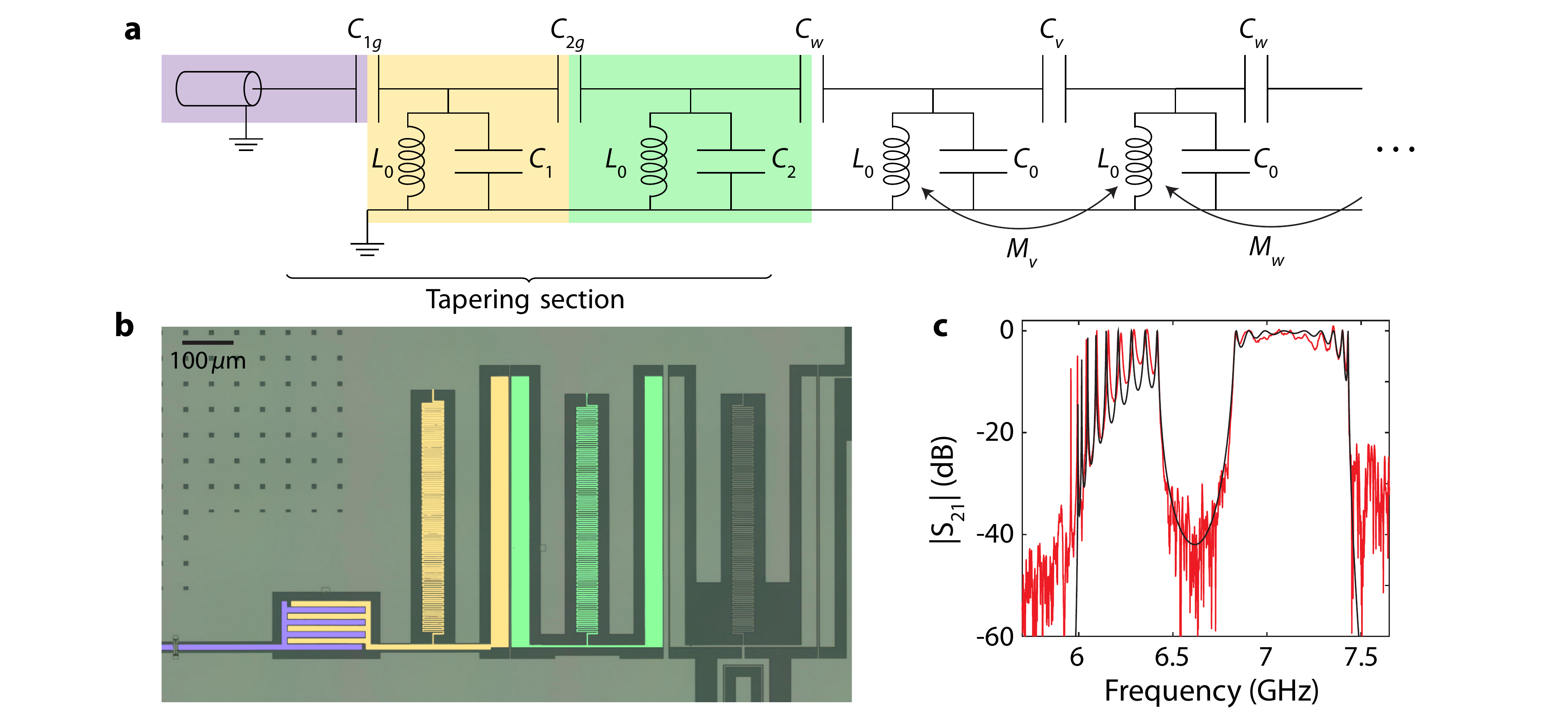}
  \caption{\textbf{Tapering section of Device I.} \textbf{a}, The circuit diagram of the tapering section connecting a coplanar waveguide to the topological waveguide. The coplanar waveguide, first tapering resonator, and second tapering resonator are shaded in purple, yellow, and green, respectively. \textbf{b,} Optical micrograph (false colored) of the tapering section on Device I. The tapering section is colored in the same manner as the corresponding components in panel \textbf{a}. \textbf{c,} Red: measured waveguide transmission spectrum $|S_{21}|$ for Device I. Black: fit to the data with parameters $(C_v, C_w) = (35, 19.2)\:\text{fF}$, $(M_v, M_w) = (-38, -32)\:\text{pH}$, $(C_{1g}, C_{2g}) = (141, 35)\:\text{fF}$, $(C_1, C_2) = (128.2, 230)\:\text{fF}$, $C_0 = 250\:\text{fF}$, $L_0 = 1.9\:\text{nH}$.}
  \label{fig:FigureS7}
\end{figure*}

\subsection{Experimental setup}
The measurement setup inside the dilution refrigerator is illustrated in Fig.~\ref{fig:FigureS6}. All the 14 qubits on Device I are DC-biased with individual flux-bias (Z control) lines, filtered by a 64\:kHz low-pass filter at the 4K plate and a 1.9 MHz low-pass filter at the mixing chamber plate. The Waveguide Input 1 (2) passes through a series of attenuators and filters including a 20 dB (30 dB) thin-film attenuator developed in B. Palmer's group~\cite{Yeh:2017}. It connects via a circulator to port 1 (2) of Device I, which is enclosed in two layers of magnetic shielding. The output signals from Device I are routed by the same circulator to the output lines containing a series of circulators and filters. The pair of 2$\times$2 switches in the amplification chain allows us to choose the branch to be further amplified in the first stage by a traveling-wave parametric amplifier (TWPA) from MIT Lincoln Laboratories. Both of the output lines are amplified by an individual high electron mobility transistor (HEMT) at the 4K plate, followed by room-temperature amplifiers at 300\:K.  All four S-parameters $S_{ij}$ ($i,j\in\{1,2\}$) involving port 1 and 2 on Device I can be measured with this setup by choosing one of the waveguide input ports and one of the waveguide output ports, e.g. $S_{11}$ can be measured by sending the input signal into Waveguide Input 1 and collecting the output signal from Waveguide Output 2 with both 2$\times$2 switches in the cross ($\times$) configuration.

\section{Tapering sections on Device I}
\label{sec:AppendixD}
The finite system size of metamaterial waveguide gives rise to sharp resonances inside the passband associated with reflection at the boundary (Fig.~\ref{fig:Figure1}d of the main text). Also, the decay rate of qubits to external ports inside the middle bandgap (MBG) is small, making the spectroscopic measurement of qubits inside the MBG hard to achieve. In order to reduce ripples in transmission spectrum inside the upper passband and increase the decay rates of qubits to external ports comparable to their intrinsic contributions inside the middle bandgap, we added two resonators at each end of the metamaterial waveguide in Device I as tapering section.

Similar to the procedure described in Appendix C of Ref.~\cite{Ferreira:2020}, the idea is to increase the coupling capacitance gradually across the two resonators while keeping the resonator frequency the same as other resonators by changing the self capacitance as well. However, unlike the simple case of an array of LC resonators with uniform coupling capacitance, the SSH waveguide consists of alternating coupling capacitance between neighboring resonators and two separate passbands form as a result. In this particular work, the passband experiments are designed to take place at the upper passband frequencies and hence we have slightly modified the resonant frequencies of tapering resonators to perform impedance-matching inside the upper passband. The circuit diagram shown in Fig.~\ref{fig:FigureS7}a was used to model the tapering section in our system. While designing of tapering sections involves empirical trials, microwave filter design software, e.g.~iFilter module in AWR Microwave Office~\cite{iFilter-MicrowaveOffice}, can be used to aid the choice of circuit parameters and optimization method.

\begin{figure*}[t!]
\centering
\includegraphics[width = 0.9\textwidth]{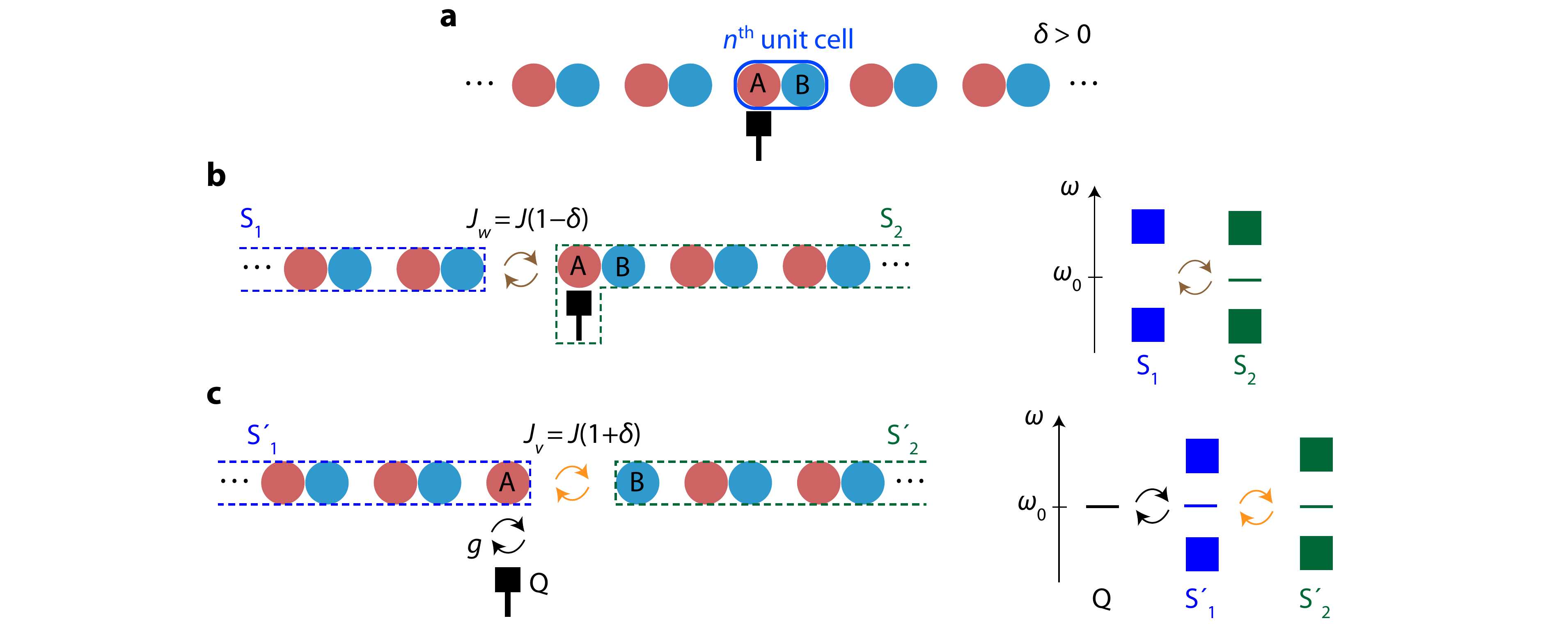}
  \caption{\textbf{Understanding the directionality of qubit-photon bound states.} \textbf{a}, Schematic of the full system consisting of an infinite SSH waveguide with a qubit coupled to the A sublattice of the $n$-th unit cell and tuned to frequency $\omega_0$ in the center of the MBG. Here we make the unit cell choice in which the waveguide is in the trivial phase ($\delta>0$).  \textbf{b}, Division of system in panel \textbf{a} into two subsystems S$_1$ and S$_2$ in Description I.  \textbf{c}, Division of system in panel \textbf{a} into three subsystems [qubit (Q), S$_1^\prime$, S$_2^\prime$] in Description II. For panels \textbf{b} and \textbf{c}, the left side shows the schematic of the division into subsystems and the right side illustrates the mode spectrum of the subsystems and the coupling between them.}
  \label{fig:FigureS8}
\end{figure*}

Figure~\ref{fig:FigureS7}b shows the optical micrograph of a tapering section on Device I. The circuit parameters are extracted by fitting the normalized waveguide transmission spectrum ($S_{21}$) data from measurement with theoretical circuit models. We find a good agreement in the frequency of normal modes and the level of ripples between the theoretical model and the experiment as illustrated in Fig.~\ref{fig:FigureS7}c. The level of ripples in the transmission spectrum of the entire upper passband is about 8\:dB and decreases to below 2\:dB near the center of the band, allowing us to probe the cooperative interaction between qubits at these frequencies.

\section{Directional shape of qubit-photon bound state} 
\label{sec:AppendixE}
In this section, we provide detailed explanations on the directional shape of qubit-photon bound states discussed in the main text. As an example, we consider a system consisting of a topological waveguide in the trivial phase and a qubit coupled to the A sublattice of the $n$-th unit cell (Fig.~\ref{fig:FigureS8}a). Our descriptions are based on partitioning the system into subsystems under two alternative pictures (Fig.~\ref{fig:FigureS8}b,c), where the array is divided on the left (Description I) or the right (Description II) of the site $(n,\text{A})$ where the qubit is coupled to.

\subsection{Description I}
We divide the array into two parts by breaking the inter-cell coupling $J_w=J(1-\delta)$ that exists on the left of the site $(n,\text{A})$ where the qubit is coupled to, i.e., between sites $(n-1,\text{B})$ and $(n,\text{A})$. The system is described in terms of two subsystems S$_1$ and S$_2$ as shown in Fig.~\ref{fig:FigureS8}b. The subsystem S$_1$ is a semi-infinite array in the trivial phase extended from the $(n-1)$-th unit cell to the left and the subsystem S$_2$ comprising a qubit and a semi-infinite array in the trivial phase extended from the $n$-th unit cell to the right. The coupling between the two subsystems is interpreted to take place at a boundary site with coupling strength $J_w$. When the qubit frequency is resonant to the reference frequency $\omega_0$, the subsystem S$_2$ can be viewed as a semi-infinite array in the topological phase, where the qubit effectively acts as an edge site. Here, the resulting topological edge mode of subsystem S$_2$ is the qubit-photon bound state, with photon occupation mostly on the qubit itself and on every B site with a decaying envelope. Coupling of subsystem S$_2$ to S$_1$ only has a minor effect on the edge mode of S$_2$ as the modes in subsystem S$_1$ are concentrated at passband frequencies, far-detuned from $\omega=\omega_0$. Also, the presence of an edge state of S$_2$ at $\omega=\omega_0$ cannot induce an additional occupation on S$_1$ by this coupling in a way that resembles an edge state since the edge mode of S$_2$ does not occupy sites on the A sublattice. The passband modes S$_1$ and S$_2$ near-resonantly couple to each other, whose net effect is redistribution of modes within the passband frequencies. Therefore, the qubit-photon bound state can be viewed as a topological edge mode for subsystem S$_2$ which is unperturbed by coupling to subsystem S$_1$. The directionality and photon occupation distribution along the resonator chain of the qubit-photon bound state can be naturally explained according to this picture.

\begin{figure*}[htb!]
\centering
\includegraphics[width = 0.9\textwidth]{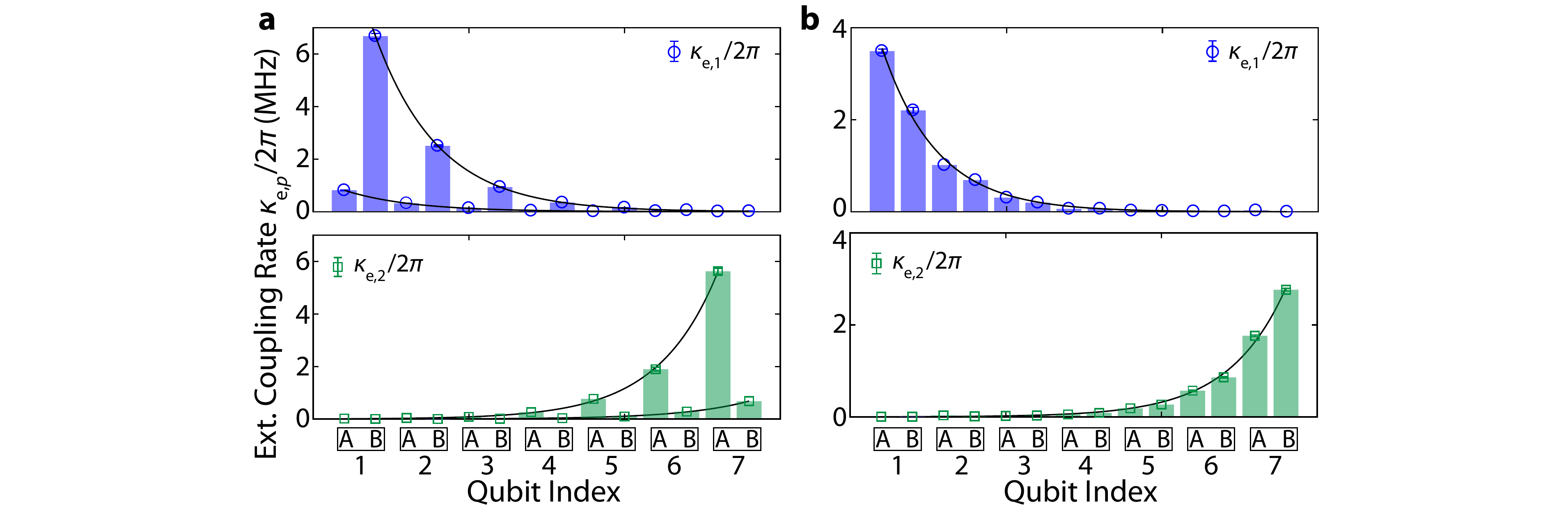}
  \caption{\textbf{Breakdown of directionality at different frequencies.} \textbf{a}, Upper (Lower) plots: external coupling rate of the qubit-photon bound states to port 1 (2) at 6.72 GHz in the middle bandgap. Exponential fit (black curve) on the data gives the localization length of $\xi=2$. \textbf{b, } Upper (Lower) plots: external coupling rate of the qubit-photon bound states to port 1 (2) at 7.485 GHz in the upper bandgap. Exponential fit (black curve) on the data gives the localization length of $\xi=1.8$. The localization lengths are represented in units of lattice constant. For all panels, the error bars show $95\%$ confidence interval and are removed on data points whose error is smaller than the marker size.}
  \label{fig:FigureS9}
\end{figure*}

\subsection{Description II}
In this alternate description, we divide the array into two parts by breaking the intra-cell coupling $J_v=J(1+\delta)$ that exists on the right of the site $(n,\text{A})$ where the qubit is coupled to, i.e., between sites $(n,\text{A})$ and $(n,\text{B})$. We consider the division of the system into three parts: the qubit, subsystem S$_1^\prime$, and subsystem S$_2^\prime$ as illustrated in Fig.~\ref{fig:FigureS8}c. Here, the subsystem S$_1^\prime$ (S$_2^\prime$) is a semi-infinite array in the topological phase extended to the left (right), where the last site hosting the topological edge mode E$_1^\prime$ (E$_2^\prime$) at $\omega=\omega_0$ is the A (B) sublattice of the $n$-th unit cell. The subsystem S$_1^\prime$ is coupled to both the qubit and the subsystem S$_2^\prime$ with coupling strength $g$ and $J_v=J(1+\delta)$, respectively. Similar to Description I, the result of coupling between subsystem modes inside the passband is the reorganization of modes without significant change in the spectrum inside the middle bandgap. On the other hand, modes of the subsystems at $\omega=\omega_0$ (qubit, E$_1^\prime$, and E$_2^\prime$) can be viewed as emitters coupled in a linear chain configuration, whose eigenfrequencies and corresponding eigenstates in the single-excitation manifold are given by
\begin{align*}
\tilde{\omega}_\pm &= \omega_0\pm \sqrt{\tilde{g}^2 + \tilde{J}_v^2},\\
|\psi_\pm\rangle &= \frac{1}{\sqrt{2}} \left(\frac{\tilde{g}}{\sqrt{\tilde{g}^2 + \tilde{J}_v^2}}|100\rangle \pm |010\rangle +\frac{\tilde{J}_v}{\sqrt{\tilde{g}^2 + \tilde{J}_v^2}} |001\rangle  \right),
\end{align*}
and
$$
\tilde{\omega}_0 = \omega_0,\quad |\psi_0\rangle = \frac{1}{\sqrt{\tilde{g}^2 + \tilde{J}_v^2}} \left(\tilde{J}_v|100\rangle - \tilde{g}|001\rangle \right),
$$
where $|n_1 n_2 n_3\rangle$ denotes a state with $(n_1, n_2, n_3)$  photons in the (qubit, E$_1^\prime$, E$_2^\prime$), respectively. Here, $\tilde{g}$ ($\tilde{J}_v$) is the coupling between edge mode E$_1^\prime$ and the qubit (edge mode E$_2^\prime$), diluted from $g$ ($J_v$) due to the admixture of photonic occupation on sites other than the boundary in the edge modes. Note that in the limit of short localization length, we recover $\tilde{g}\approx g$ and $\tilde{J}_v\approx J_v$. Among the three single-excitation eigenstates, the states $|\psi_\pm\rangle$ lie at frequencies of approximately $\omega_0 \pm J$, and are absorbed into the passbands. The only remaining state inside the middle bandgap is the state $|\psi_0\rangle$, existing exactly at $\omega=\omega_0$, which is an anti-symmetric superposition of qubit excited state and the single-photon state of E$_2^\prime$, whose photonic envelope is directed to the right with occupation on every B site. This accounts for the directional qubit-photon bound state emerging in this scenario. 

\section{Coupling of qubit-photon bound states to external ports at different frequencies}
\label{sec:AppendixF}
As noted in the main text (Fig.~\ref{fig:Figure2}), the perfect directionality of the qubit-photon bound states is achieved only at the reference frequency $\omega_0$ inside the middle bandgap. In this section, we discuss the breakdown of the observed perfect directionality when qubits are tuned to different frequencies inside the middle bandgap by showing the behavior of the external coupling $\kappa_{\text{e},p}$ ($p=1,2$) to the ports.

\subsection{Inside the middle bandgap, detuned from the reference frequency}
Figure~\ref{fig:FigureS9}a shows the external coupling rate of qubits to the ports at 6.72\:GHz, a frequency in the middle bandgap close to band-edge. The alternating behavior of external coupling rate is still observed, but with a smaller contrast than in Fig.~\ref{fig:Figure2} of the main text. The dependence of external linewidth on qubit index still exhibits the remaining directionality with qubits on A (B) sublattice maintaining large coupling to port 2 (1), while showing small non-zero coupling to the opposite port.

\begin{figure*}[tb!]
\centering
\includegraphics[width=\textwidth]{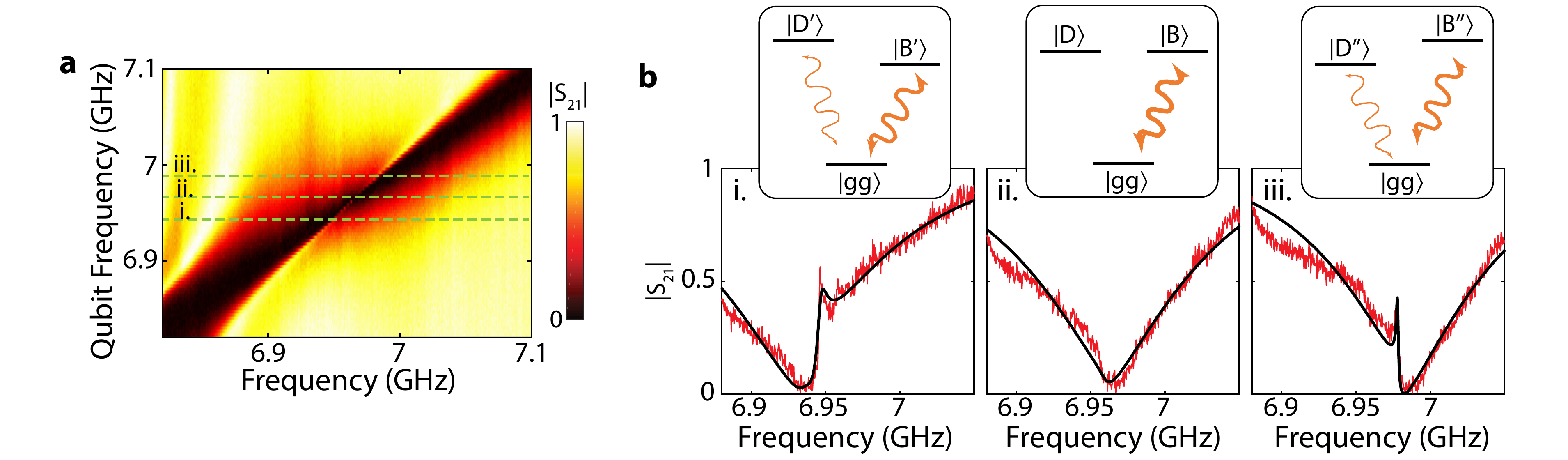}
  \caption{\textbf{Bright and dark states of a qubit pair coupled to a topological waveguide.} \textbf{a}, Zoomed-in view of the swirl feature near 6.95\:GHz of the experimental data illustrated in Fig.~\ref{fig:Figure3}c in the main text. \textbf{b,} Transmission spectrum across two-qubit resonance for three different frequency tunings, corresponding to line cuts marked with green dashed lines on panel \textbf{a}. The insets to panel \textbf{b} show the corresponding level diagram with $|\text{gg}\rangle$ denoting both qubits in ground states and $|\text{B}\rangle$ ($|\text{D}\rangle$) representing the perfect bright (dark) state. The state notation with prime (double prime) in sub-panel {i.}~({iii.})~denotes the imperfect super-radiant bright state and sub-radiant dark state, with the width of orange arrows specifying the strength of the coupling of states to the waveguide channel. The sub-panel {ii.}~occurs at the center of the swirl, where perfect super-radiance and sub-radiance takes place (i.e., bright state waveguide coupling is maximum and dark state waveguide coupling is zero). The black and red curves correspond to experimental data and theoretical fit, respectively.}
  \label{fig:FigureS10}
\end{figure*}

\subsection{Inside the upper bandgap}
Inside the upper bandgap (7.485\:GHz), the coupling of qubit-photon bound states to external ports decreases monotonically with the distance of the qubit site to the port, regardless of which sublattice the qubit is coupled to (Fig.~\ref{fig:FigureS9}b). This behavior is similar to that of qubit-photon bound states formed in a structure with uniform coupling, where bound states exhibit a symmetric photonic envelope surrounding the qubit. Note that we find the external coupling to port 2 ($\kappa_{\text{e},2}$) to be generally smaller than that to port 1  ($\kappa_{\text{e},1}$), which may arise from a slight impedance mismatch on the connection of the device to the external wiring.

\section{Probing band topology with qubits}
\label{sec:AppendixG}
\subsection{Signature of perfect super-radiance}

Here we take a closer look at the swirl pattern in the waveguide transmission spectrum -- a signature of perfect super-radiance -- which is discussed in Fig.~\ref{fig:Figure3}c of the main text. In Fig.~\ref{fig:FigureS10} we zoom in to the observed swirl pattern near 6.95\:GHz, and three horizontal line cuts. At the center of this pattern (sub-panel ii.~of Fig.~\ref{fig:FigureS10}b), the two qubits form perfect super-/sub-radiant states with maximized correlated decay and zero coherent exchange interaction~\cite{vanLoo:2013,Lalumiere:2013}. At this point, the transmission spectrum shows a single Lorentzian lineshape (perfect super-radiant state and bright state) with linewidth equal to the sum of individual linewidths of the coupled qubits. The perfect sub-radiant state (dark state), which has no external coupling, cannot be accessed from the waveguide channel here and is absent in the spectrum. Slightly away from this frequency, the coherent exchange interaction starts to show up, making hybridized states $|\text{B}^\prime\rangle$, $|\text{D}^\prime\rangle$ formed by the interaction of the two qubits. In this case, both of the hybridized states have non-zero decay rate to the waveguide, forming a V-type level structure~\cite{Bello:2019}. The interference between photons scattering off the two hybridized states gives rise to the peak in the middle of sub-panels (i.) and (iii.) in Fig.~\ref{fig:FigureS10}b. 

The fitting of lineshapes starts with the subtraction of transmission spectrum of the background, which are taken in the same frequency window but with qubits detuned away. Note that the background subtraction in this case cannot be perfect due to the frequency shift of the upper passband modes under the presence of qubits. Such imperfection accounts for most of the discrepancy between the fit and the experimental data. The fit employs the transfer matrix method discussed in Refs.~\cite{loo2014interactions, shen2005coherenta, shen2005coherentb}. Here, the transfer matrix of the two qubits takes into account the pure dephasing, which causes the sharp peaks in sub-panels (i.) and (iii.) of Fig.~\ref{fig:FigureS10}b to stay below perfect transmission level (unity) as opposed to the prediction from the ideal case of electromagnetically induced transparency~\cite{Witthaut:2010}. 

\begin{figure*}[ht!]
\centering
\includegraphics[width = \textwidth]{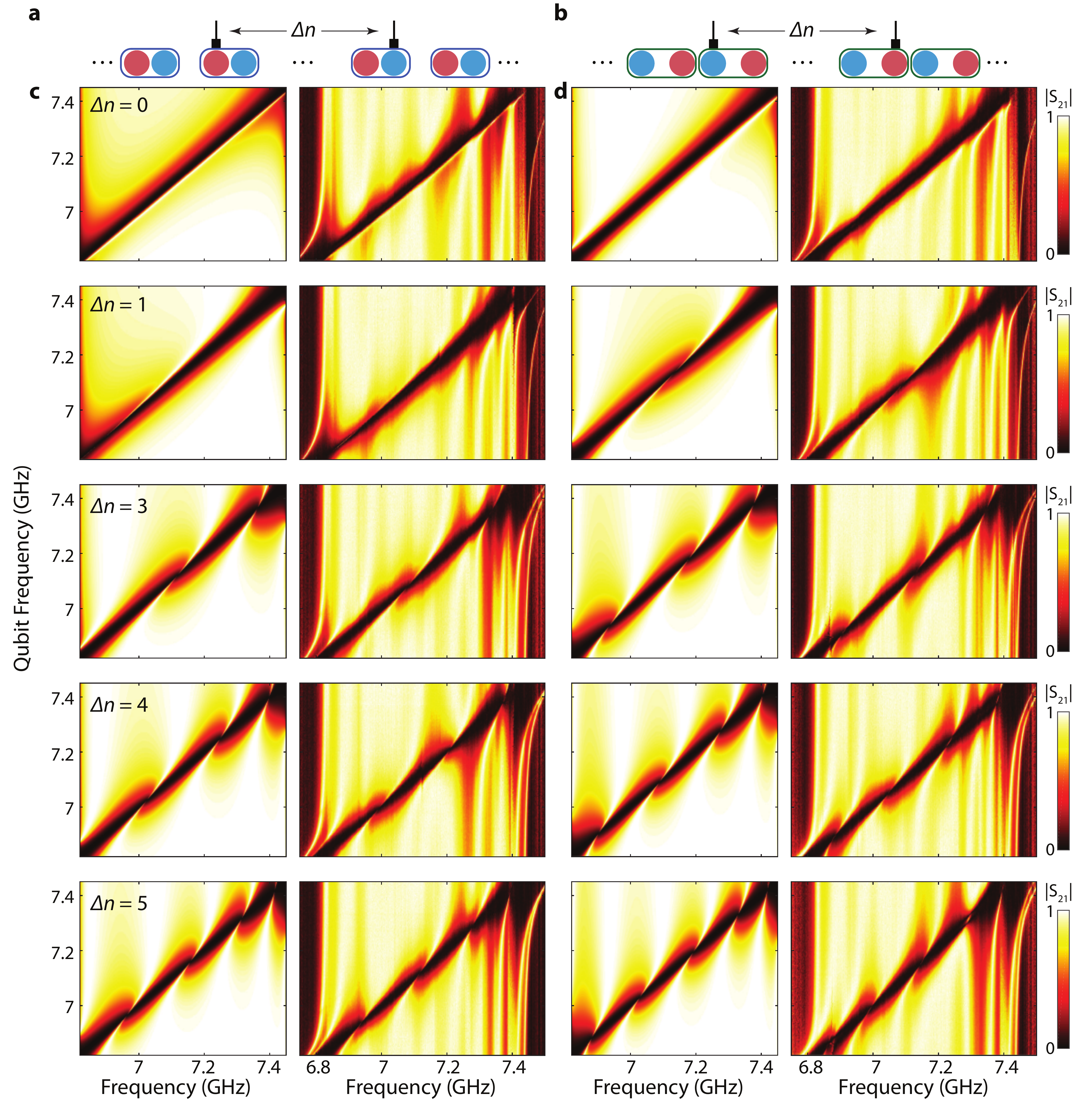}
  \caption{\textbf{Probing the passband topology with qubit pairs.} \textbf{a}, Schematic showing two qubits separated by $\Delta n$ unit cells in the trivial configuration. \textbf{b}, Corresponding schematic for topological phase configuration.  \textbf{c}, Waveguide transmission spectrum $|S_{21}|$ when frequencies of two qubits are resonantly tuned across the upper passband in the trivial configuration.  \textbf{d}, Waveguide transmission spectrum $|S_{21}|$ for the topological configuration.  For both trivial and topological spectra, the left spectrum illustrates theoretical expectations based on Ref.~\cite{Bello:2019} whereas the right shows the experimental data.}
  \label{fig:FigureS11}
\end{figure*}

\subsection{Topology-dependent photon scattering on various qubit pairs}
As mentioned in the main text, when the two qubits are separated by $\Delta n$ ($\Delta n > 0$) unit cells, perfect super-radiance (vanishing of coherent exchange interaction) takes place exactly $\Delta n -1$ times in the trivial phase and $\Delta n$ times in the topological phase across the entire passband. The main text shows the case of $\Delta n = 2$. Here we report similar measurements on other qubit pairs with different cell distance $\Delta n$ between the qubits. Figure~\ref{fig:FigureS11} shows good qualitative agreement between the experiment and theoretical result in Ref.~\cite{Bello:2019}. The small avoided-crossing-like features in the experimental data are due to coupling of one of the qubits with a local two-level system defect. An example of this is seen near 6.85\:GHz of $\Delta n = 3$ in the topological configuration. For $\Delta n = 0$, there is no perfect super-radiant point throughout the passband for both trivial and topological configurations. For all the other combinations in Fig.~\ref{fig:FigureS11}, the number of swirl patterns indicating perfect super-radiance agrees with the theoretical model.

\begin{figure*}[tb!]
\centering
\includegraphics[width = \textwidth]{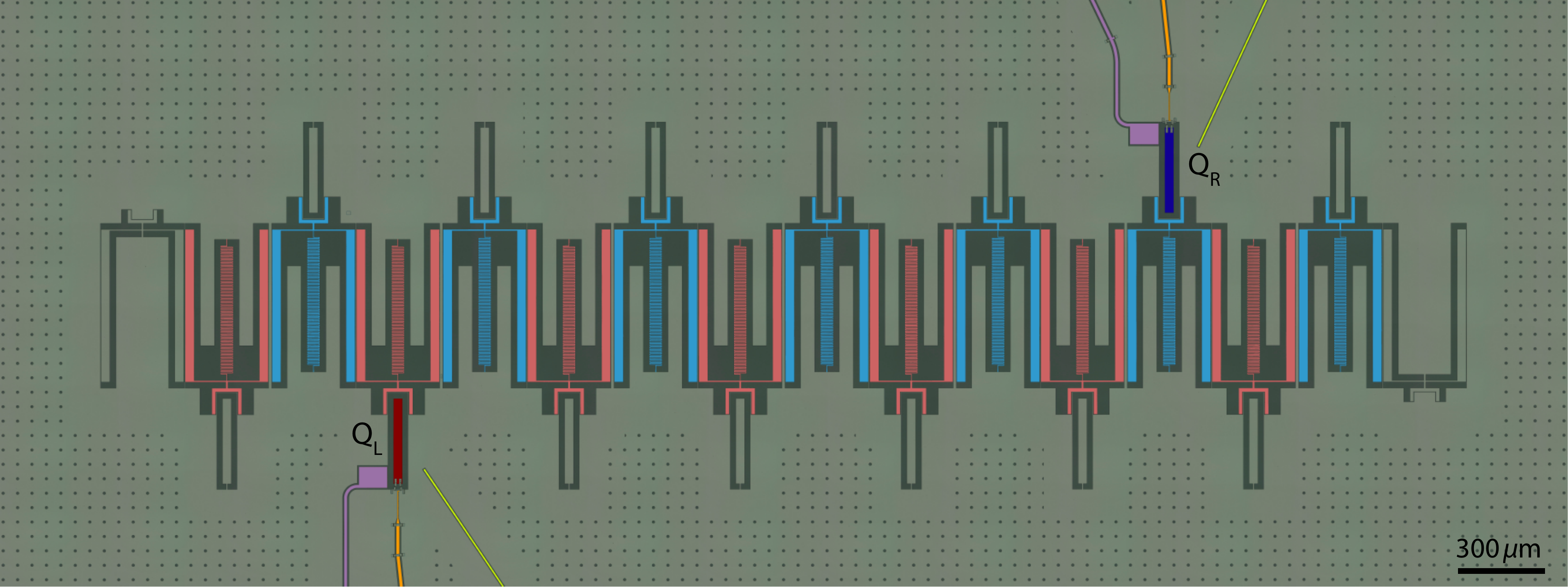}
  \caption{\textbf{Optical micrograph of Device II (false-colored).} The device consists of a topological waveguide with 7 unit cells (resonators corresponding to A/B sublattice colored red/blue) in the topological phase, where the inter-cell coupling is larger than the intra-cell coupling. Two qubits Q$_\text{L}$ (dark red) and Q$_\text{R}$ (dark blue) are coupled to A sublattice of the second unit cell and B sublattice of sixth unit cell, respectively. Each qubit is coupled to a $\lambda/4$ coplanar waveguide resonator (purple) for dispersive readout, flux-bias line (orange) for frequency control, and charge line (yellow) for local excitation control.
  }
  \label{fig:FigureS12}
\end{figure*}

\begin{table*}[htb!]
\begin{threeparttable}[t]
  \centering
\begin{tabular}{M{1.0cm}P{1.3cm}P{1.3cm}P{1.3cm}P{1.3cm}P{1.3cm}P{1.3cm}P{1.0cm}P{1.0cm}} 
              Qubit & $f_{\text{max}}$ (GHz)   & $E_C/h$ (MHz)    & $E_{J\Sigma}/h$ (GHz)    & $g_E/2\pi$ (MHz)  & $f_{\text{RO}}$ (GHz)   & $g_{\text{RO}}/2\pi$ (MHz)   & $T_1$ ($\mu$s) & $T_2^*$ ($\mu$s)  \\ \hline \hline
$\text{Q}_\text{L}$   &    8.23 & 294 & 30.89 & 58.1 & 5.30 & 43.5 & 4.73 & 4.04 \\
$\text{Q}_\text{R}$   &    7.99 & 296 & 28.98 & 57.3 & 5.39 & 43.4 & 13.9 & 8.3  
 \end{tabular}
   \end{threeparttable}
 \caption{\textbf{Qubit parameters on Device II.} $f_{\text{max}}$ is the maximum frequency (sweet spot) and $E_C$ ($E_{J\Sigma}$) is the charging (Josephson) energy of the qubit. $g_E$ is the coupling of qubit to the corresponding edge state. The readout resonator at frequency $f_{\text{RO}}$ is coupled to the qubit with coupling strength $g_{\text{RO}}$. $T_1$ ($T_2^*$) is the lifetime (Ramsey coherence time) of a qubit measured at the sweet spot.}
   \label{tb:qubitsII}
\end{table*}

\section{Device II characterization and experimental setup}
\label{sec:AppendixH}
In this section we provide a detailed description of the elements making up Device II, in which the edge mode experiments are performed. The optical micrograph of Device II is illustrated in Fig.~\ref{fig:FigureS12}.

\subsection{Qubits}
The parameters of qubits on Device II are summarized in Table~\ref{tb:qubitsII}. The two qubits are designed to have identical SQUID loops with symmetric JJs. The lifetime and Ramsey coherence times in the table are measured when qubits are tuned to their sweet spot. Qubit coherence at the working frequency in the middle bandgap is also characterized, with the lifetime and Ramsey coherence times of $\text{Q}_\text{L}$ ($\text{Q}_\text{R}$) at 6.829 (6.835)\:GHz measured to be $T_1=6.435$ (5.803)\:$\mu$s and $T_2^*=344$ (539)\:ns, respectively.

\subsection{Metamaterial waveguide and coupling to qubits}
The resonators in the metamaterial waveguide and their coupling to qubits are designed to be nominally identical to those in Device I. The last resonators of the array are terminated with a wing-shape patterned ground plane region in order to maintain the bare self-capacitance identical to other resonators.

\subsection{Edge modes}
The coherence of the edge modes is characterized by using qubits to control and measure the excitation with single-photon precision. Taking $\text{E}_\text{L}$ as an example, we define the iSWAP gate as a half-cycle of the vacuum Rabi oscillation in Fig.~\ref{fig:Figure4}d of the main text.
For measurement of the lifetime of the edge state E$_\text{L}$, the qubit Q$_\text{L}$ is initially prepared in its excited state with a microwave $\pi$-pulse, and an iSWAP gate is applied to transfer the population from $\text{Q}_\text{L}$ to $\text{E}_\text{L}$. After waiting for a variable delay, we perform the second iSWAP to retrieve the population from $\text{E}_\text{L}$ back to $\text{Q}_\text{L}$, followed by the readout of $\text{Q}_\text{L}$. In order to measure the Ramsey coherence time, the qubit Q$_\text{L}$ is instead prepared in an equal superposition of ground and excited states with a microwave $\pi/2$-pulse, followed by an iSWAP gate. After a variable delay, we perform the second iSWAP and another $\pi/2$-pulse on Q$_\text{L}$, followed by the readout of $\text{Q}_\text{L}$. An equivalent pulse sequence for Q$_\text{R}$ is used to characterize the coherence of E$_\text{R}$. The lifetime and Ramsey coherence time of $\text{E}_\text{L}$ ($\text{E}_\text{R}$) are extracted to be $T_1=3.68$ (2.96)\:$\mu$s and $T_2^*=4.08$ (2.91)\:$\mu$s, respectively, when $\text{Q}_\text{L}$ ($\text{Q}_\text{R}$) is parked at 6.829 (6.835)\:GHz. Due to the considerable amount of coupling $g_E$ between the qubit and the edge mode compared to the detuning at park frequency, the edge modes are hybridized with the qubits during the delay time in the above-mentioned pulse sequences.  As a result, the measured coherence time of the edge modes is likely limited here by the dephasing of the qubits.

\begin{figure*}[htb!]
\centering
\includegraphics[width = \textwidth]{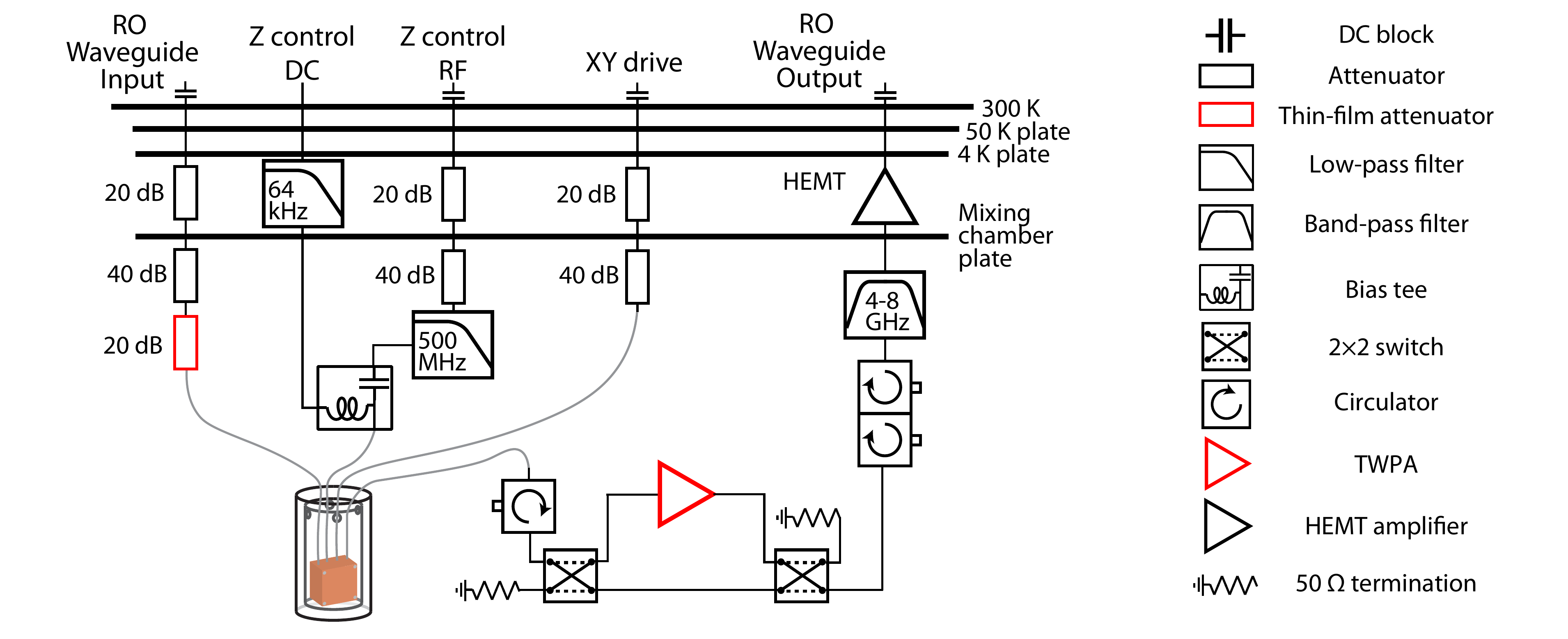}
  \caption{\textbf{Schematic of the measurement setup inside the dilution refrigerator for Device II.} The meaning of each symbol in the schematic on the left is enumerated on the right. The level of attenuation of each attenuator is indicated with number next to the symbol. The cutoff frequencies of each filter is specified with numbers inside the symbol. Small squares attached to circulator symbols indicate port termination with $Z_0=50\:\Omega$, allowing us to use the 3-port circulator as a 2-port isolator. The pump line for the TWPA is not shown in the diagram for simplicity.}
  \label{fig:FigureS13}
\end{figure*}

\subsection{Experimental setup}
The measurement setup inside the dilution refrigerator is illustrated in Fig.~\ref{fig:FigureS13}. The excitation of the two qubits is controlled by capacitively-coupled individual XY microwave drive lines. The frequency of qubits are controlled by individual DC bias (Z control DC) and RF signals (Z control RF), which are combined using a bias tee at the mixing chamber plate. The readout signals are sent into RO Waveguide Input, passing through a series of attenuators including a 20 dB thin-film attenuator developed in B.~Palmer’s group~\cite{Yeh:2017}. The output signals go through an optional TWPA, a series of circulators and a band-pass filter, which are then amplified by a HEMT amplifier (RO Waveguide Output).

\subsection{Details on the population transfer process}
In step i) of the double-modulation scheme described in the main text, the frequency modulation pulse on $\text{Q}_\text{R}$ (control modulation) is set to be 2\:ns longer than that on $\text{Q}_\text{L}$ (transfer modulation). The interaction strength induced by the control modulation is 21.1 MHz, smaller than that induced by the transfer modulation in order to decrease the population leakage between the two edge states. For step iii), the interaction strength induced by the control modulation on $\text{Q}_\text{L}$ is 22.4 MHz, much closer to interaction strength for the transfer than expected (this was due to a poor calibration of the modulation efficiency of qubit sideband). The interaction strengths being too close between $\text{Q}_\text{L} \leftrightarrow \text{E}_\text{L}$ and $\text{Q}_\text{R} \leftrightarrow \text{E}_\text{R}$ gives rise to unwanted leakage and decreases the required interaction time in step ii). We expect that a careful optimization on the frequency modulation pulses would have better addressed this leakage problem and increase the transfer fidelity (see below).

The fit to the curves in Fig.~\ref{fig:Figure4}e of the main text are based on numerical simulation with QuTiP~\cite{Johansson:2012,Johansson:2013}, assuming the values of lifetime ($T_1$) and coherence time ($T_2^*$) from the characterization measurements. The free parameters in the simulation are the coupling strengths $\Tilde{g}_\text{L}$, $\Tilde{g}_\text{R}$ between qubits and edge states, whose values are extracted from the best fit of the experimental data. 

The detailed contributions to the infidelity of the as-implemented population transfer protocol are also analyzed by utilizing QuTiP. The initial left-side qubit population probability is measured to be only 98.4\:\%, corresponding to an infidelity of 1.6\:\% in the $\pi$-pulse qubit excitation in this transfer experiment (compared to a previously calibrated `optimized' pulse). In the following steps, we remove the leakage between edge modes and the decoherence process sequentially to see their individual contributions to infidelity. First, we set the coupling strength between the two edge modes to zero during the two iSWAP gates while keeping the above-mentioned initial population probability, coupling strengths, lifetimes, and coherence times. The elimination of unintended leakage during the left and right side iSWAP steps between the edge modes gives the final transferred population probability of 91.9\:\%, suggesting $91.9\:\% - 87\:\% = 4.9\:\%$ of the infidelity comes from the unintended leakage between edge modes. Also, as expected, setting the population decay and decoherence of the qubits and the edge modes to zero, the final population is found to be identical to the initial value, indicating that $98.4\:\% - 91.9\:\% = 6.5\:\%$ of loss arises from the decoherence processes. 


%

\end{document}